\documentclass[aps,prd,xsuperscriptaddress,nofootinbib,longbibliography,floatfix]{revtex4-2}

\usepackage[T1]{fontenc}
\usepackage[utf8]{inputenc} 
\usepackage{microtype}
\usepackage{amsmath,amssymb,amsfonts}
\usepackage[mathlines]{lineno}
\usepackage{aas_macros}
\usepackage{bm}
\usepackage{graphicx}
\usepackage{dcolumn}
\usepackage{booktabs}
\usepackage{multirow}
\usepackage{orcidlink}

\usepackage{soul}
\usepackage{xcolor}

\newcommand{\rchi}{\mathord{\mathchoice
{\raisebox{0.15ex}{$\displaystyle\chi$}}
{\raisebox{0.15ex}{$\textstyle\chi$}}
{\raisebox{0.10ex}{$\scriptstyle\chi$}}
{\raisebox{0.08ex}{$\scriptscriptstyle\chi$}}}}

\usepackage{hyperref} 
\hypersetup{ 
colorlinks=true, 
linkcolor=blue,
citecolor=magenta,
urlcolor=magenta,
pdfpagemode=UseOutlines, 
pdfstartview=FitH, 
bookmarksopen=true
}

\begin{document}


\title{Transition from Diffusion to Drift-Dominated Cosmic Ray Transport and the Origin of the Knee}

\author{Luis E. Espinosa Castro\orcidlink{0009-0004-6030-2788}} 
\email{luis.espinosacastro@gssi.it}
\affiliation{Gran Sasso Science Institute (GSSI), Viale Francesco Crispi 7, 67100 L'Aquila, Italy}
\affiliation{INFN-Laboratori Nazionali del Gran Sasso (LNGS),  via G. Acitelli 22, 67100 Assergi (AQ), Italy}

\author{Carmelo Evoli\orcidlink{0000-0002-6023-5253}} 
\email{carmelo.evoli@gssi.it}
\affiliation{Gran Sasso Science Institute (GSSI), Viale Francesco Crispi 7, 67100 L'Aquila, Italy}
\affiliation{INFN-Laboratori Nazionali del Gran Sasso (LNGS),  via G. Acitelli 22, 67100 Assergi (AQ), Italy}

\author{Pasquale Blasi\orcidlink{0000-0003-2480-599X}}
\email{pasquale.blasi@gssi.it}
\affiliation{Gran Sasso Science Institute (GSSI), Viale Francesco Crispi 7, 67100 L'Aquila, Italy}
\affiliation{INFN-Laboratori Nazionali del Gran Sasso (LNGS),  via G. Acitelli 22, 67100 Assergi (AQ), Italy}

\date{\today}

\begin{abstract}
{In a magnetic field with a complex topology, as can be the Galactic magnetic field, cosmic ray transport cannot simply be described by diffusion parallel and perpendicular to magnetic field lines, because the gradients and curvature of the large-scale magnetic field induce drift motions. These effects become especially important at high energies. Here we revisit the possibility that the competition between diffusion and drifts may lead to a \emph{knee} in the cosmic ray spectrum. We carry out test-particle simulations of cosmic ray transport in a mock Galactic magnetic field made of a regular large scale component, with a non-trivial topology and a homogeneous and isotropic turbulent magnetic field, with a spectrum that is assumed to be Kolmogorov-like in the basic setup. These simulations are used to infer the escape time and the grammage accumulated by cosmic rays with energy in the TeV--10 PeV energy range. In the case of a large scale magnetic field with a purely azimuthal structure, the drift due to the curvature of magnetic field lines produces a \emph{knee} in the PeV range, but the model fails to reproduce the grammage, due to the exceedingly low value of the perpendicular diffusion coefficient. If the large scale magnetic field acquires a component perpendicular to the Galactic disc, the parallel diffusion coefficient becomes quickly dominant in terms of particle escape, and drifts are unable to compete. A \emph{knee} structure does not appear in such a scenario. However, if the parallel diffusion coefficient becomes energy independent at $E\gtrsim 1$ TeV, a \emph{knee} may arise around PeV energies due to drift dominance. We discuss two cases in which this situation may occur.}
\end{abstract}

\maketitle

\section{Introduction}
\label{sec:introduction}

The \emph{knee} is one of the most prominent features in the cosmic-ray (CR) all-particle spectrum measured at Earth. Since its discovery, it has been interpreted as a key signature of the transition between different regimes of Galactic CR acceleration and transport. 
In its simplest phenomenological description, the all-particle spectrum steepens from approximately $E^{-2.7}$ to $E^{-3.0}$--$E^{-3.1}$ at an energy of a few PeV. However, the physical interpretation of this feature remains unsettled, because the observed all-particle spectrum results from the superposition of different nuclear species.

A crucial piece of information comes from direct measurements of CR nuclei below the knee. Space-borne and balloon-borne experiments have shown that the spectra of individual elements are not simple power laws: protons and helium display different spectral shapes, primary nuclei harden at rigidities of a few hundred GV, and both proton and helium spectra show evidence for further softening in the multi-TeV range~
\cite{Adriani2011,Yoon2011,Aguilar2015,Yoon2017,An2019,Aguilar2021,Alemanno2021,Adriani2022,Adriani2023}. These results do not directly determine the knee, because of the limited exposure of direct experiments at PeV energies, but they provide the indispensable low-energy boundary condition for any model that attempts to extrapolate CR spectra into the knee region.

The experimental determination of the knee itself has therefore relied mainly on air-shower measurements. A major step was provided by KASCADE, which showed that the first knee is associated primarily with the steepening of the light component rather than with a simultaneous break of all nuclear species~\cite{Antoni2005}. 
KASCADE-Grande then extended this picture by reporting a knee-like structure in the heavy component around $8\times 10^{16}\,\mathrm{eV}$ and an ankle-like hardening in the light component around $10^{17.1}\,\mathrm{eV}$~\cite{Apel2011,Apel2013}.
Other air-shower, Cherenkov, fluorescence, and hybrid measurements, including Tibet-III, IceTop/IceCube, Tunka-133, and TALE, have established that the region between the knee and the ankle is not featureless, but contains a sequence of structures that are naturally discussed in terms of rigidity-dependent mass groups~\cite{Amenomori2008,Aartsen2019,Budnev2020,Abbasi2018}.

Recently, the Large High Altitude Air Shower Observatory (LHAASO) published measurements of the CR all-particle spectrum and mean logarithmic mass across the \emph{knee}~\cite{Cao2024all} in the energy range $0.3$--$30$ PeV. The all-particle knee is found at $3.67 \pm 0.05 \pm 0.15$ PeV, with a change in spectral index from $\gamma \simeq -2.74$ to $\gamma \simeq -3.13$. The mean logarithmic mass decreases up to $\sim 3$ PeV and then shows a trend towards heavier composition above the knee. The fact that the knee position and the change in power-law index are approximately aligned with the behaviour of the light component suggests that the all-particle knee is closely connected to the steepening of light nuclei.

The same collaboration has recently released measurements of the proton and helium spectra across the knee region~\cite{Cao2025protons,Cao2026helium}. The proton spectrum shows a sharp softening around a rigidity of a few PV, while the helium spectrum displays a more complex behaviour, with a hardening and a subsequent softening at higher energy. These new measurements provided additional information: a successful model should not only reproduce the position of the all-particle knee, but also account for the rigidity ordering, the size of the spectral steepening, the evolution of the composition, and, ideally, the behaviour of the large-scale anisotropy.

The origin of the \emph{knee} remains elusive. A broad class of models associates it with the maximum rigidity of particles accelerated in Galactic sources. In this picture, the all-particle knee reflects the progressive cutoff of individual nuclear species, following the classic Peters cycle~\cite{Peters1961,Horandel2004}. 
Such models naturally predict a gradual increase of the mean mass above the proton knee, but they require Galactic accelerators to reach maximum rigidities of order $\sim \mathrm{PV}$. This requirement is demanding for standard supernova remnants (SNRs), unless magnetic-field amplification and favourable environmental conditions are invoked~\cite{Cristofari2020}. For this reason, other classes of Galactic accelerators, such as young stellar clusters and microquasars, have also been proposed as possible contributors to the PeV domain~\cite{Aharonian2019,Kaci2025,Zhang2026}.

A second class of models attributes the knee to propagation rather than to the maximum energy of the sources. In this case, the observed steepening is produced by a change in the rigidity dependence of the escape time from the Galaxy. Such models have the attractive feature that they introduce a natural scale: in a $\sim\mu{\rm G}$ Galactic magnetic field, the Larmor radius of a PeV proton is of order a parsec, comparable to characteristic scales of interstellar magnetic turbulence. This observation has a long history, going back to early arguments that magnetic confinement in the Galaxy should become progressively inefficient at sufficiently high rigidity~\cite{Cocconi1956}.
At the same time, they impose an even stronger requirement on acceleration, because the maximum rigidity of the sources must then extend appreciably beyond the observed knee. One possibility is that the knee marks the transition from resonant diffusion to the small-pitch-angle scattering regime, which occurs when the particle Larmor radius becomes comparable to or larger than the coherence length of the turbulent magnetic field. In this regime, the diffusion coefficient scales approximately as $D(E)\propto E^2$~\cite{DeMarco2007,Giacinti2015,Subedi2017,Dundovic2020}. 
The corresponding steepening of individual mass components, $\Delta\gamma \sim 1.5$--$1.7$, is however significantly larger than suggested by the recent LHAASO measurements of the light components.

Another way in which transport can shape the knee is through the onset of drift motions. This idea was first proposed in Ref.~\cite{Ptuskin1993}, where the knee was interpreted as the result of a transition from diffusion-dominated escape to drift-dominated escape in the large-scale Galactic magnetic field (GMF). In a magnetized plasma, gradients and curvature of the regular field generate an antisymmetric contribution to the diffusion tensor, usually described by the Hall diffusion coefficient~\cite{Isenberg1979}. Since the corresponding drift velocity grows approximately linearly with rigidity, drifts may become competitive with perpendicular diffusion at sufficiently high energy. In the estimates of Ref.~\cite{Ptuskin1993}, this transition can occur at a few PeV and produces a spectral steepening of order $\Delta\gamma\sim 0.7$, close to the observed size of the knee. The scenario was further developed in Refs.~\cite{Candia2002,Candia2004,Evoli2007}, where test-particle simulations were used to estimate the energy dependence of diffusion and drift coefficients and to connect the first knee, the second knee, and the expected anisotropy.

Despite its appeal, the drift interpretation has not yet been tested under conditions that simultaneously reproduce the relevant Galactic confinement time, grammage, and magnetic-field geometry. This point is important because the observational constraints on CR transport are not limited to the PeV region. At lower energies, secondary-to-primary ratios and radioactive clocks constrain the grammage and residence time of CRs in the Galaxy, while at higher energies the anisotropy and the second-knee region provide additional tests. A viable drift model must therefore be assessed not only by asking whether it can generate a knee, but also by asking whether the same transport setup remains compatible with the broader phenomenology of Galactic CR propagation.

In this article, we investigate the onset of drifts and their interplay with both parallel and perpendicular diffusion in shaping the CR spectrum at very high energies. To this end, we perform test-particle simulations of charged particles propagating in synthetic turbulent magnetic fields superimposed on an axisymmetric coherent magnetic field that mimics the large-scale GMF. We consider configurations with and without a component perpendicular to the Galactic disk, in order to clarify how the geometry of the ordered field controls the dominant escape channel.

We first compute the diffusion coefficients parallel and perpendicular to the large-scale magnetic field, establishing a direct connection with previous numerical studies. We then determine the drift contribution and compare it with theoretical expectations. We show that, for a purely azimuthal ordered magnetic field, a knee-like feature can arise from the competition between perpendicular diffusion and drift escape. However, the corresponding confinement time and accumulated grammage are difficult to reconcile with the extrapolation to lower energies, where secondary-to-primary ratios and unstable isotopes constrain CR transport. If a sizeable vertical component of the large-scale field is introduced, parallel escape becomes faster than drift escape and the knee-like transition disappears. Finally, we show that if the parallel mean free path becomes approximately energy independent above TeV energies, drifts can again become competitive with diffusion and produce a knee at roughly the observed energy without violating grammage constraints. We discuss possible physical situations in which such an energy-independent parallel mean free path may arise.

This work is organized as follows. In Sec.~\ref{sec:theory}, we recall the analytical description of CR diffusion and drifts. In Sec.~\ref{sec:simulations}, we describe the numerical setup of our test-particle simulations. The results are presented in Sec.~\ref{sec:results}. Finally, in Sec.~\ref{sec:conclusion}, we discuss the implications of our findings for CR transport and assess the viability of antisymmetric drift as an explanation of the CR knee.

\section{Diffusion-Drift Transition}
\label{sec:theory}

Before presenting the numerical test-particle simulations, it is useful to recall the transport framework that motivates our study. At TeV--PeV energies, Galactic CR propagation is mainly controlled by their interaction with the large-scale and turbulent components of the Galactic magnetic field. In this regime, spatial transport is not necessarily described by diffusion alone: gradients and curvature of the regular magnetic field generate an antisymmetric, or drift, contribution to the transport tensor. This term becomes increasingly important with rigidity and can therefore introduce a characteristic energy scale in the propagated spectrum. The purpose of this section is to outline this diffusion--drift transition and to show why drift effects provide a natural mechanism to investigate in connection with the CR knee~\cite{Ptuskin1993,Candia2002}. Although the results presented in Section~\ref{sec:results} are obtained with a test-particle approach, the standard transport-equation formulation recalled below provides the physical basis for interpreting the simulations and identifying the relevant scaling with energy and magnetic-field geometry.

At these energies, energy losses and spallation can be safely neglected, so that the steady-state transport equation for CRs propagating in a turbulent GMF is given by
\begin{equation}
\label{eq:transport_equation}
\nabla \cdot \textbf{J} = - \nabla_i D_{ij}(\textbf{r}) \nabla_j N(\textbf{r}) = Q(\textbf{r}),
\end{equation}
where $\textbf{J}$ is the CR current, $N(\textbf{r})$ is the CR density, $Q(\textbf{r})$ is the source term, and $D_{ij}(\textbf{r})$ is the diffusion tensor.

The most general form of this tensor can be written in terms of three diffusion coefficients: the parallel coefficient $D_\parallel$, defined with respect to the direction of the regular guiding field; the perpendicular coefficient $D_\perp$; and the antisymmetric, or Hall, coefficient $D_A$~(e.g.,~\cite{Giacalone1999}). 
Explicitly,
\begin{equation}
\label{eq:diffusion_tensor}
D_{ij}= (D_\parallel - D_\perp)b_ib_j + D_\perp\delta_{ij} - D_A\epsilon_{ijk}b_k,
\end{equation}
where $b_i=B_i/B_0$ are the components of the unit vector along the direction of the regular magnetic field, $\delta_{ij}$ is the Kronecker delta and $\epsilon_{ijk}$ is the Levi-Civita tensor.
The parallel diffusion coefficient describes the random motion of the particles due to resonant interactions with fluctuations in the magnetic field on small scales, comparable with the particle gyroradius~\cite{Jokipii1966}. The perpendicular diffusion coefficient is instead dominated by the large scale fluctuations of the field and describes diffusive motion perpendicular to the regular field~\cite{Blasi:2023quf}. The Hall diffusion coefficient is related to the macroscopic motion of the CR distribution, and as one can infer from its contribution to the transport equation (see~Eq.~\ref{eq:azimuthal_equation}), it describes drifts~\cite{Isenberg1979}.

As a first instance, we assume that the large scale GMF is practically toroidal ($B_\phi$ is the dominant component), meaning that the field lines follow closed circles. In this case we can rewrite Eq.~\ref{eq:transport_equation} in cylindrical coordinates as
\begin{equation}
\label{eq:azimuthal_equation}
    \left(-\frac{1}{r}\frac{\partial}{\partial r}\left[rD_\perp\frac{\partial}{\partial r}\right] - \frac{\partial}{\partial z}\left[D_\perp\frac{\partial}{\partial z}\right] + u_r\frac{\partial}{\partial r} + u_z\frac{\partial}{\partial z}
    \right) N(\textbf{r}) = Q(\textbf{r}),
\end{equation}
where we have used the azimuthal symmetry in our system ($b_r=b_z=0$, $b_\phi=\pm1$) to remove all derivatives with respect to $\phi$. For Eq.~\ref{eq:azimuthal_equation}, we have used the divergence of the CR current in cylindrical coordinates, written as
\begin{equation}
\label{eq:current}
    \nabla \cdot \textbf{J} = \frac{1}{r}\frac{\partial}{\partial r}(rJ_r) + \frac{1}{r}\frac{\partial}{\partial \phi}J_\phi + \frac{\partial}{\partial z}J_z.
\end{equation}

The left-hand side of Eq.~\ref{eq:azimuthal_equation} contains two diffusion terms, proportional to the perpendicular component of the diffusion tensor, and two additional terms that can be interpreted as drift contributions.
The corresponding drift velocities are proportional to the Hall coefficient and are given by
\begin{eqnarray}
\label{eq:drift1}
u_r & = & -\frac{\partial (D_A b_\phi)}{\partial z} \\
\label{eq:drift2}
u_z & = & \frac{1}{r}\frac{\partial (rD_A b_\phi)}{\partial r}
\end{eqnarray}
These expressions make explicit the role of the antisymmetric part of the diffusion tensor, which is directly responsible for CR drift motions. Importantly, such drifts can arise even in the absence of spatial gradients in the strength of the large-scale magnetic field, provided that its direction varies in space. More generally, drift effects are expected whenever the magnetic-field gradients or curvatures occur on scales larger than the particle Larmor radius $r_L(E)$~\cite{Isenberg1979,Tautz2012},
\begin{equation}
\label{eq:drift_inequality}
r_L=\frac{p}{ZeB} \ll L=\left|\frac{1}{B}\frac{\partial B_i}{\partial x_j}\right|^{-1}.
\end{equation}
The resulting drift motion is perpendicular to both the coherent magnetic field and the direction associated with its gradient or curvature. Its velocity can be written as~\cite{Rossi1970}
\begin{equation}
\label{eq:drift_complete}
u_\perp = c,r_L \left[\frac{1}{2}\sin^2{\alpha}\frac{\vec{B_0}\times\nabla B_0}{B_0^2} + \cos^2{\alpha}\frac{\vec{B_0}\times((\vec{B_0}\cdot\nabla)\vec{B_0})}{B_0^3}\right] = c \, r_L \cos^2{\alpha}\left[\frac{(\nabla\times\vec{B_0})_\perp}{B_0}\right],
\end{equation}
where $c$ is the speed of light and $\alpha$ is the particle pitch angle. This expression is consistent with the drift terms appearing in the transport-equation formulation and shows explicitly that the drift velocity is proportional to the Larmor radius. Since $r_L$ increases with particle rigidity, drift effects are expected to become increasingly relevant at high energies, as we discuss in the following.

In this configuration, the effect of drifts on the escape of CRs from the Galactic halo can be understood by comparing the drift escape timescale, $\tau_{\rm drift}=H/u_\perp$, with the diffusive escape timescale from a halo of size $H$. For the large-scale magnetic configuration considered here, only perpendicular diffusion contributes to vertical escape from the halo, so that $\tau_{\rm diff} \sim H^2/D_\perp$.

Because diffusion and drift have different energy dependences, one naturally expects a transition from diffusion-dominated escape at low energies to drift-dominated escape at high energies. For a Kolmogorov spectrum of interstellar turbulence, the perpendicular diffusion coefficient in the resonant regime scales as $D_\perp \sim E^{1/2}$~\cite{Dundovic2020}, whereas drift motions are approximately linear in energy. The transition between the two regimes would therefore produce a spectral steepening of order $\Delta\gamma \sim 0.5$, and could occur in the PeV range for plausible Galactic parameters, as originally suggested in Refs.~\cite{Ptuskin1993,Candia2002} as a possible explanation of the \emph{knee}.

This picture differs from other propagation-motivated interpretations of the \emph{knee}. For instance, in Ref.~\cite{Giacinti2015}, the authors proposed that the \emph{knee} may originate from a change in the \emph{parallel} diffusion regime, from resonant scattering, $D_\parallel \sim E^{1/3}$, to small-pitch-angle scattering, $D_\parallel \sim E^2$, occurring when the Larmor radius becomes comparable to the coherence length of the turbulent magnetic field. In that scenario, the expected spectral break is $\Delta \gamma \gtrsim 1$, which is in tension with current measurements of the \emph{knee} in individual mass components~\cite{Cao2025protons,Cao2026helium}.

It is important to stress that these different regimes apply to different large-scale magnetic-field configurations. Perpendicular diffusion is relevant for vertical escape when the regular field is close to purely azimuthal, whereas parallel diffusion requires a sizeable component of the large-scale magnetic field perpendicular to the Galactic disk. The geometry of the regular Galactic magnetic field therefore plays a central role in determining which transport regime controls CR escape, and hence in assessing whether a propagation-induced transition can account for the observed properties of the \emph{knee}.

We highlight that all models in which the \emph{knee} is associated to a change in the properties of particle transport require that the maximum energy of the accelerated particles exceeds the one corresponding to the transition to drift dominated propagation and the energy where $r_L\sim \lambda_c$. Both these requirements are very demanding in terms of acceleration theory, given the daunting difficulties faced by models even in reaching PeV energies~\cite{Ptuskin2010,Schure2013,Cristofari2020,Caprioli:2023orv}. 

\section{Test-Particle Simulations}
\label{sec:simulations}

We perform test particle simulations in synthetic turbulent magnetic fields with the Monte Carlo code CRPropa 3.2\footnote{\href{https://crpropa.desy.de/}{crpropa.desy.de}}~\cite{AlvesBatista2022}. 
In this section, we present an overview of our numerical setups. We describe the magnetic field configurations and the details regarding our computation of CR diffusion coefficients (explained in subsection~\ref{sec:sim_1}) and the CR escape times from the Galaxy and the grammage accumulated in their trajectories (described in subsection~\ref{sec:sim_2}).

\par For all our simulations, the magnetic field is composed of a turbulent component over a coherent field with strength $B_0=1$ $\mu$G. The turbulent magnetic field is constructed via a 3D grid in Fourier space, i.e.~$B_i (\vec k)$, with wavenumber $k$. The field components in $\vec x$-space are computed by inverse Fast Fourier Transform on the grid. The magnitudes of the field at each grid vertex are sampled from a Gaussian distribution (with \textit{root mean square} field strength $\sqrt{\langle\delta B^2\rangle} = \eta \, B_0$, for a turbulence level $\eta\lesssim 1$) and follow an isotropic turbulence spectrum. Following Ref.~\cite{Dundovic2020}, this spectrum can be written as:
\begin{equation}
\label{eq:spectrum}
    W(k) = \frac{C(q,s)}{\pi k^2}\delta B^2 l_b\frac{(kl_b)^q}{[1 + (kl_b)^2]^{(s+q)/2}} 
\end{equation}
for a bend-over scale $l_b$, over a wavenumber range limited by the minimum and maximum turbulent scales ($k_\text{min} = 1/L_{\text{max}}$ and $k_{\text{max}} = 1/L_{\text{min}}$, respectively).
We adopt an isotropic Kolmogorov spectrum, for which $s=5/3$ and $q=4 $. 
The normalization $C(q,s)$ is expressed as
\begin{equation}
\label{eq:spectrum_normalization}
    C(q,s) = \frac{\Gamma(\frac{s+q}{2})}{2\Gamma(\frac{s-1}{2})\Gamma(\frac{q+1}{2})}, 
\end{equation}
with $\Gamma(x)$ the Euler gamma function. 

We use a $2048^3$ grid, with spacing $\Delta = 0.015$ pc between grid points. This grid is periodically extended to fill the simulation box. To maximize the diffusion resonant range, the minimum turbulence scale is set to $L_{\text{min}} = 2 \, \Delta$, while the maximum scale is $L_{\text{max}} = (N_{\text{grid}} \, \Delta) / 2$. 
\par For an isotropic turbulent field, the correlation length is directly proportional to the bend-over scale and can be computed, as in Ref.~\cite{Dundovic2020}, by
\begin{equation}
\label{eq:correlation_length}
    \lambda_c = \frac{4\pi}{\delta B^2}\int_0^\infty dr \int_0^\infty dk  \frac{\sin(kr)}{kr} k^2 W(k) = \frac{4\pi}{s(s+2)}C(q,s)l_b.
\end{equation}
For our simulations, we assume a bend-over scale of $l_b\approx60$ pc, corresponding to a correlation length of the order of $10$ pc, as suggested by different interstellar tracers~\cite{Haverkorn2008,Iacobelli2013}.

\subsection{Components of the diffusion tensor}
\label{sec:sim_1}

For this initial set of simulations, the coherent magnetic field was set as uniform along the x-direction (i.e. $\vec{B_0} = B_0\hat{x}$), to simplify the measurement of the diffusion coefficients in Eq.~\ref{eq:diffusion_tensor}.

The calculations presented in this section are mainly aimed at establishing contact with the existing literature, in that the calculation of the parallel and perpendicular diffusion coefficients has been presented elsewhere \cite{Casse2001,DeMarco2007,Tautz2012,Dundovic2020,Mertsch2025a,Mertsch2025b}. 
For the Hall diffusion coefficient, we follow Refs.~\cite{Tautz2012,Candia2004}, adopting it as an effective description of gradient and curvature drifts whenever the magnetic field is spatially inhomogeneous or locally curved.
Notice that the Hall diffusion coefficient is non-zero even when the magnetic field is homogeneous, but its interpretation as a drift is related to the gradients through
\begin{equation}
\langle v_{D,i}\rangle = \frac{\partial D_{ij}}{\partial x_j}. 
\end{equation}
Hence, if the antisymmetric components of the diffusion tensor are spatially independent (as one would expect if the turbulence is homogeneous and the regular field is spatially constant), the drift velocity vanishes. 

The positions and directions of simulated particles with equal rigidity are tracked at time intervals $t_n = n \, 0.01 \, r_L(E)/c$, for $n\in[0,10^4]$, then the simulations are repeated for different values of rigidity, covering the range $r_L/\lambda_c\in[10^{-2}, 10]$. The diffusion coefficients are given, according to the Taylor–Green–Kubo (TGK) formula (e.g.,~\cite{Giacalone1999}), by
\begin{equation}
\label{eq:tgk}
    D_{ij} = \int_0^\infty dt R_{ij}(t) =\int_0^\infty dt \left< v_i(t) v_j(0)\right>,
\end{equation}
where $i,j \in \{x,y,z\}$ and $R_{ij}(t)$ are functions measuring the decorrelation of particle trajectories, computed from the product of the velocity components $\left< v_i(t) v_j(0)\right>$ at time $t$, where $\left<\right>$ indicates the average over the entire population of particles with the same energy. We assume the ansätze described in Ref.~\cite{Bieber1997} for these decorrelation functions, expressing them as exponentially decaying functions
\begin{equation}
\label{eq:parallel_decorrelation}
    R_\parallel(t) = R_{zz}(t) = \frac{c^2}{3}e^{-t/\tau_\parallel},
\end{equation}
\begin{equation}
\label{eq:perp_decorrelation}
    R_\perp(t) = R_{xx}(t) = R_{yy}(t) = \frac{c^2}{3}e^{-t/\tau_\perp}\cos{\omega t},
\end{equation}
\begin{equation}
\label{eq:hall_decorrelation}
    R_A(t) = - R_{xy}(t) = R_{yx}(t) =\frac{c^2}{3}e^{-t/\tau_A}\sin{\omega t}.
\end{equation}
with $\tau_\parallel$, $\tau_\perp$, and $\tau_A$ being the decorrelation timescales and $\omega=c/r_L$ is the gyro-frequency of the particles. We checked that indeed these functional forms do provide a good description of the temporal evolution of the decorrelation. Introducing these ansätze into Eq.~\ref{eq:tgk}, we obtain
\begin{equation}
\label{eq:parallel_coeffcient}
    D_\parallel = \frac{c^2 \tau_\parallel}{3},
\end{equation}
\begin{equation}
\label{eq:perp_coeffcient}
    D_\perp = \frac{c^2}{3}  \frac{\tau_\perp}{1 + (\omega\tau_\perp)^2},
\end{equation}
\begin{equation}
\label{eq:hall_coefficient}
    D_A = \frac{c^2}{3}  \frac{\omega\tau_A^2}{1 + (\omega\tau_A)^2}.
\end{equation}
We then fit the decorrelation times measured in our simulations using Eqs.~\ref{eq:parallel_decorrelation}--\ref{eq:hall_decorrelation}, obtaining the corresponding decorrelation timescales at each energy. These timescales are then inserted into Eqs.~\ref{eq:parallel_coeffcient}--\ref{eq:hall_coefficient} to derive the diffusion coefficients as functions of energy. This procedure allows us to compute all components of the diffusion tensor, including a non-vanishing antisymmetric coefficient.

We stress that fitting the decorrelation times measured in the simulations does not, by itself, provide a physical explanation for these timescales. Rather, it should be understood as an alternative method to determine the diffusion coefficients. We checked that the values of $D_\parallel$ and $D_\perp$ obtained in this way are consistent with those found in previous studies using other methods, such as the asymptotic value of the time-dependent running diffusion coefficient~(e.g.,~\cite{Giacalone1999,Dundovic2020}),
\begin{equation}
D_{ij}(t) = \frac{\left< \Delta x_i(t) \, \Delta x_j(t) \right>}{2t} \, .
\end{equation}
The advantage of the method adopted here is that it also allows us to determine the Hall diffusion coefficient with sufficiently low numerical noise.

\begin{figure}[t]
\centering
  \includegraphics[width=0.497\linewidth]{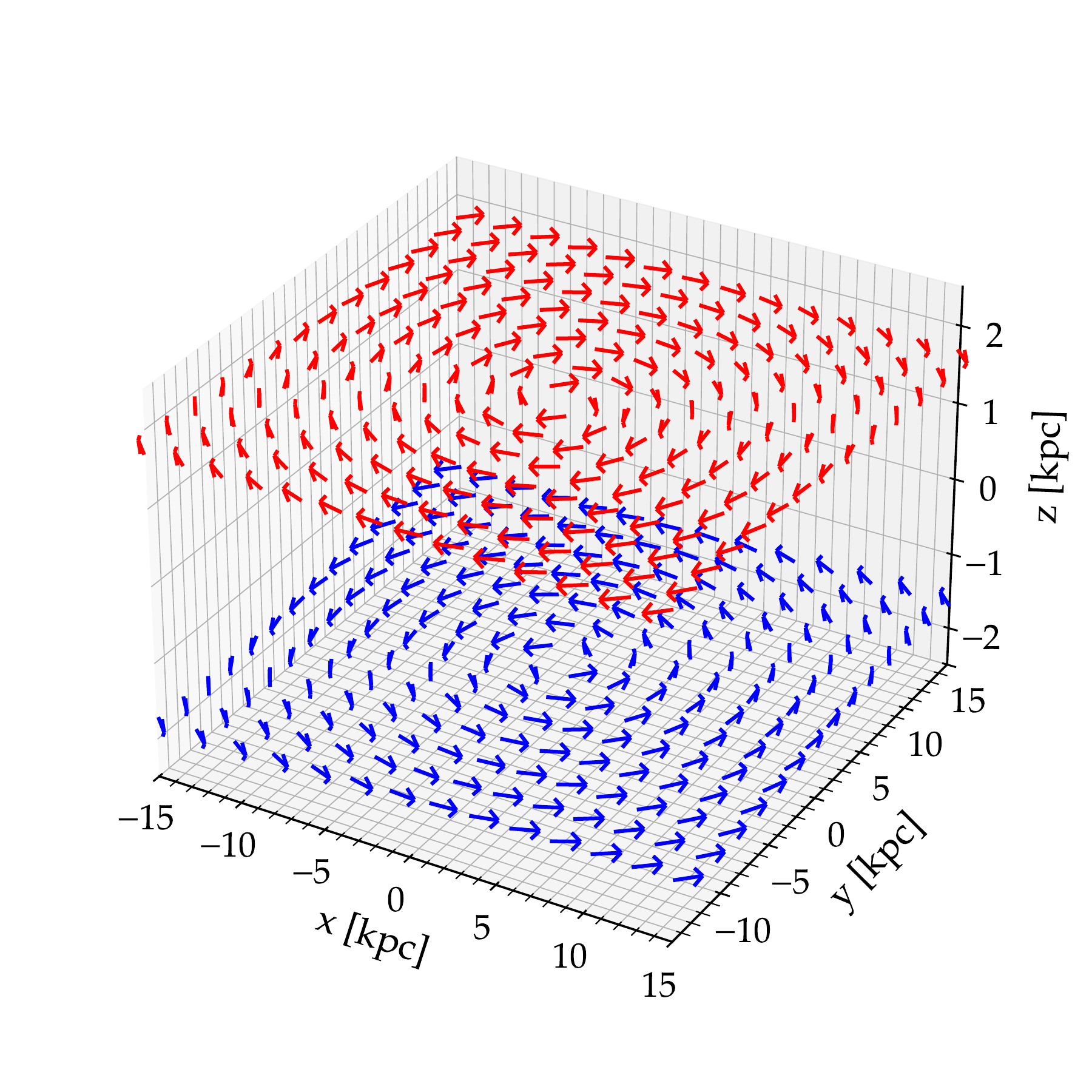}
  \includegraphics[width=0.497\linewidth]{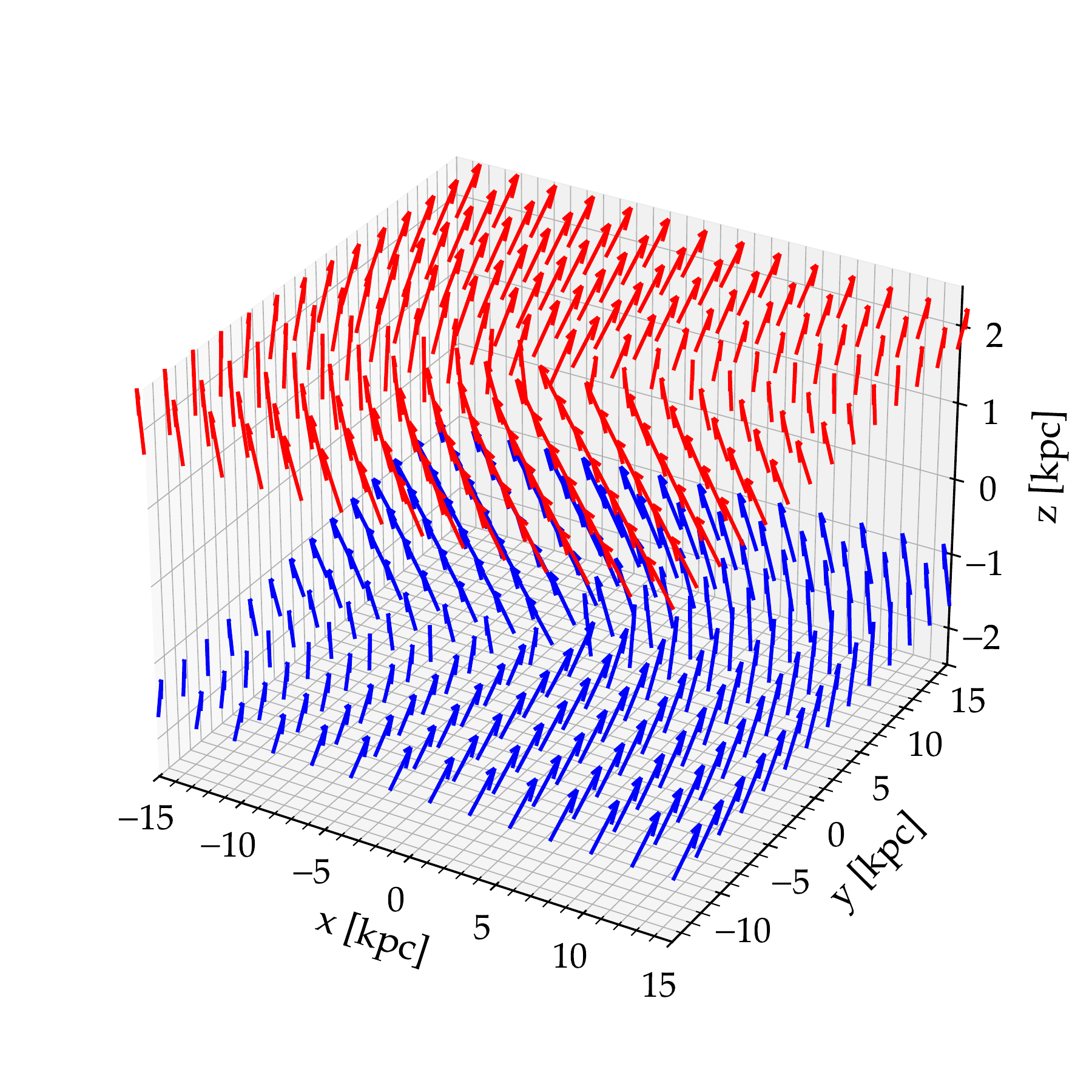}
\caption{Three-dimensional diagrams of coherent magnetic field geometry in our simulations for a fully-azimuthal antisymmetric field ($\mu=0$, left panel) and an azimuthal field with a non-zero vertical component ($\mu=0.2$, right panel). Arrows represent direction of magnetic field lines. Different arrow colors are chosen for better visualization. Grid in the diagrams do not represent the exact grid used in the simulations.}
\label{fig:diagrams_field}
\end{figure}

\subsection{Cosmic ray escape time and grammage}
\label{sec:sim_2}

The measured quantities that are typically used to infer the Galactic diffusion coefficient and in general the transport properties of Galactic CRs are the escape time and the grammage accumulated by CRs during propagation. Typically, these quantities are deduced from the measured values of the abundance of unstable isotopes (for instance through ratios such as $\rm^{10}Be/^9Be$ or Be/B) and through secondary/primary ratios of the abundances of stable nuclei. The combination of these measurements imposes strong constraints on the size of the confinement volume (halo) and the transport properties~\cite{Evoli:2023kxd}. 

Simulations such as the ones discussed here are usually a compromise between having a realistic description of CR transport, having sufficiently simple assumptions that the individual physical ingredients can be identified, and computational limitations. 

For the simulations discussed here, we focus on the propagation of cosmic rays in  the TeV-PeV energy range in a Galactic-like environment, schematized as follows: our simulation box is created considering a Galactic radius of 15 kpc and a Galactic halo height of $H=2$ kpc, a lower limit based on gamma-ray observations of high-velocity clouds by Fermi-LAT~\cite{Tibaldo2015}, but also marginally compatible with the lower limits found from $\rm Be/B$ ratio \cite{Evoli2020,Weinrich2020}.  

The coherent field is constructed to resemble the large-scale structure of the GMF in the disk and halo, which is predominantly toroidal~\cite{Ferriere2011,Unger2024,Haverkorn2026arXiv}. For this, we assume an azimuthal coherent magnetic field in Galactic plane. Additionally, we consider the possibility of a a small fraction $\mu$ of the field strength in the vertical direction. Thus, our coherent field is expressed as
\begin{equation}
\label{eq:coherent_field}
    B_0^2(x,y,z) = \begin{pmatrix}
(1-\mu)B_0^2 \, \frac{y^2}{x^2+y^2} \\[5pt]
- (1-\mu)B_0^2 \, \frac{x^2}{x^2+y^2} \\[5pt]
\mu B_0^2
\end{pmatrix}.
\end{equation}
Moreover, the magnetic field in the halo seems to show antisymmetric features, i.e., opposite directions above and below the Galactic disk, which could be attributed to a dynamo origin of the field~\cite{Han2001,Pshirkov2011,Xu2024}.
Therefore, we consider the field to be antisymmetric with respect to the Galactic plane, $B_\phi(z)=-B_\phi(-z)$.
For our base scenario, we assume a coherent field with closed field-lines by setting $\mu=0$, as originally proposed in Ref.~\cite{Ptuskin1993}. 
We later consider non-zero values of $\mu$ and evaluate their impact on the CR escape from the simulation box, as it is discussed in section~\ref{sec:results_B}.
A schematic representation of our coherent field configuration is shown in Fig.~\ref{fig:diagrams_field} for $\mu=0,0.2$. The field-line directions are represented as arrows. Two planes  of the field (with constant $z$-value) are displayed with different colors to facilitate their visualization. The left panel of Fig.~\ref{fig:diagrams_field} corresponds to the purely azimuthal antisymmetric field ($\mu=0$), while the right panel shows the tilting of field-lines as the main effect of adding a small vertical component ($\mu\neq0$).

In this setup, we inject several thousands of particles isotropically at the Sun position ($R_\odot=8.5$ kpc, $z=0$) with an energy spectrum going as $\propto E^{-1}$ (in order to have a uniform energy distribution in logarithmic scale) over the range  $r_L(E)/\lambda_c\in[10^{-2}, 10]$. 

Following Ref.~\cite{DeMarco2007}, we track the position, energy and propagation time of these particles from their injection and until they reach the halo height $|z|=H$, after which they are considered out of the simulation box and are no longer tracked. 
Then, for each simulated particle, we extract their escape time $\tau(E)$ and their grammage $X(E)$ accumulated along their total trajectories. We bin the particles over energy and take the average escape time in each bin. For the computation of the grammage, we integrate the interstellar medium (ISM) gas density over the trajectory of the particles. We assume a constant ISM number density of $0.5$ cm$^{-3}$ inside the Galactic disk (which is considered to have a thickness of $2h=100$ pc) and $10^{-3}$ cm$^{-3}$ outside ($h<|z|<H$).

It is worth stressing that the diffusion tensor discussed above is never formally used in the computation of the escape time and grammage. The tensor components are only used to provide an \emph{a posteriori} physical interpretation of the results of our simulations.
  
\section{Results}
\label{sec:results}

In this section, we present the results of our simulations in terms of parallel, perpendicular and Hall diffusion (Section~\ref{sec:results_A}), escape time of CRs and grammage for the purely azimuthal ordered magnetic field (Section~\ref{sec:results_B}) and for the case in which a small component along the $z$-axis is assumed (i.e. $\mu\neq0$ in Eq.~\ref{eq:coherent_field}). Finally, in section~\ref{sec:results_D} we discuss two models of the turbulent magnetic field that lead to energy independent parallel diffusion and discuss the physical implications of such models in terms of structure of the \emph{knee}. 

\subsection{Components of the diffusion tensor}
\label{sec:results_A}

\begin{figure}[t]
\centering
  \includegraphics[width=0.5\linewidth]{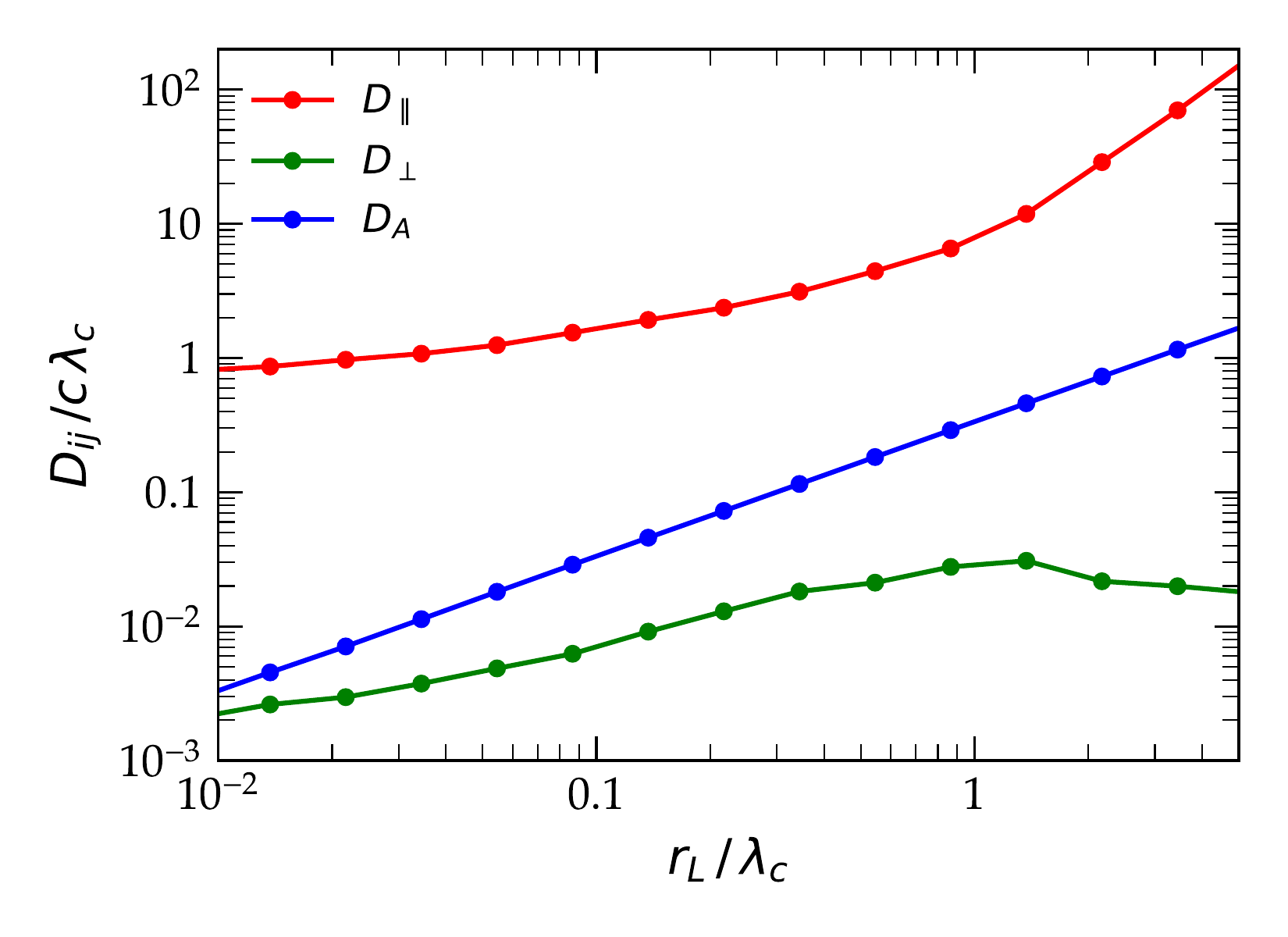}
\caption{Parallel (red), perpendicular (green) and antisymmetric (blue) coefficients of the diffusion tensor (Eq.~\ref{eq:diffusion_tensor}) as function of particle rigidity for a turbulence amplitude $\delta B / B_0=0.5$, extracted from our simulations. Scatter points represent average diffusion coefficient over rigidity bin. Coefficients normalized over $c\lambda_c$ and particle rigidity represented by ratio $r_L(E)/\lambda_c$.}
\label{fig:coefficients}
\end{figure}

In Fig.~\ref{fig:coefficients} we show the parallel (red), perpendicular (green) and Hall (blue) diffusion coefficients as functions of particle rigidity, for turbulence level $\eta=0.5$. The rigidity has been expressed here in dimensionless form through the ratio of the Larmor radius and the coherence length, $r_L(E)/\lambda_c$. We confirm previous findings that the energy dependence of $D_\parallel$ for $r_L(E)/\lambda_c\lesssim 1$ is the same that would have been derived naively from quasi-linear theory (QLT, $D_\parallel\propto E^{1/3}$). At $r_L(E)/\lambda_c\gg 1$, the universal scaling $D_\parallel\propto E^{2}$ is recovered, independent of the type of turbulence and the value of $\eta$ (see, for instance, \cite{Subedi2017} in the case of fully turbulent magnetic field). 

In the range of $r_L(E)/\lambda_c$ accessible to our simulations, we find that $D_\perp(E)\propto E^{0.5}$, confirming a result first found by~\cite{DeMarco2007} and now confirmed by all numerical simulations of transport in synthetic spatially homogeneous and isotropic turbulence~\cite{Dundovic2020}. 
However, recently it has become clear that for $r_L(E)/\lambda_c\lesssim 10^{-2}$ one recovers the expectation that the ratio $D_\perp/D_\parallel$ becomes independent of energy \cite{Mertsch2025a,Mertsch2025b}. 

We confirm that the antisymmetric coefficient $D_A\propto r_L$ over the entire accessible rigidity range, as predicted in Ref.~\cite{Ptuskin1993}. Notice that this would translate to an energy dependent drift only in the presence of a large scale gradient of the magnetic field, as we discuss below.  

\subsection{Galactic escape of cosmic rays in azimuthal coherent magnetic field ($\mu=0$)}
\label{sec:results_B}

When the magnetic field is assumed to be purely azimuthal, particles can escape the propagation volume only through perpendicular diffusion and drifts. In this case, the drifts are due to the curvature of the magnetic field lines and act in the direction perpendicular to the disc (namely along $z$). The timescales for escape due to perpendicular diffusion and drifts can easily be estimated, and it is instructive to do so. The perpendicular diffusion coefficient can be estimated from a heuristic argument put forward by \cite{Shalchi2019}, in which the running perpendicular diffusion coefficient is connected with the running parallel diffusion coefficient and with the diffusion coefficient of magnetic field lines:
\begin{equation}
 d_{\perp}(t)=\frac{d_{\rm FL}(s)}{s}\,d_{\parallel}(t),
\end{equation}
where $s=\sqrt{2d_{\parallel}t}$. Following \cite{Shalchi2019}, we assume that this result remains valid up until a time $\tau_c$ when the covered perpendicular distance becomes of order $\lambda_c$. At that point $D_\perp\simeq \lambda_c^2/2\tau_c$. Following this procedure one can obtain a rough estimate of the perpendicular diffusion coefficient as $D_\perp (E)\approx \eta^4 D_\parallel (E)$. Clearly this approach does not catch the phenomenon described above that causes a non constant ratio $D_\perp/D_\parallel$ for $10^{-2}\lesssim r_L\,\lambda_c\lesssim 1$, but for the sake of a simple estimate this is not too important, provided we keep in mind all of its limitations. The time scale for perpendicular escape can now be written as 
\begin{equation}
    \tau_{d,\perp}\sim \frac{H^2}{D_\perp}\approx \frac{H^2}{\eta^4 D_\parallel}\approx \frac{3 H^2}{\eta^2}\frac{1}{c \lambda_c}\left(\frac{r_L}{\lambda_c} \right)^{-1/3},
\end{equation}
where we have assumed for $D_\parallel$ an expression inspired by QLT: $D_\parallel\approx (1/3)c \lambda_c \eta^{-2} (r_L/\lambda_c)^{1/3}$, specific to the case of Kolmogorov turbulence. On the other hand, if one uses Eq.~\ref{eq:drift_complete} for the drift velocity, averaged over the pitch angle, one gets for the drift escape time:
\begin{equation}
    \tau_{dr,\perp}\sim \frac{H}{v_\perp}\approx \frac{H}{\frac{1}{2}c r_L/L}\approx \frac{2 H L}{\lambda_c c}\left( \frac{r_L}{\lambda_c}\right)^{-1},
\end{equation}
where we assumed that the curvature acts on a scale length $L$. Requiring that escape may be dominated by drifts implies, neglecting terms of order unity, the following condition:
\begin{equation}
    \frac{r_L}{\lambda_c} \gtrsim \left( \frac{L}{H}\right)^{3/2} \eta^3.
\end{equation}
For the numerical values of the parameters that we chose it is clear that the transition to drift dominated transport may easily occur for $r_L/\lambda_c<1$, a necessary condition to use the approximate expression for $D_\parallel$ introduced earlier, and values of the energy around $\sim$ PeV.

It is equally simple to see that a similar transition occurs on the grammage, but the low energy extrapolation of the grammage cannot be made compatible with the one inferred from Galactic observations of the secondary/primary ratios, as discussed below. 

\begin{figure}[t]
    \centering
    \includegraphics[width=0.55\linewidth]{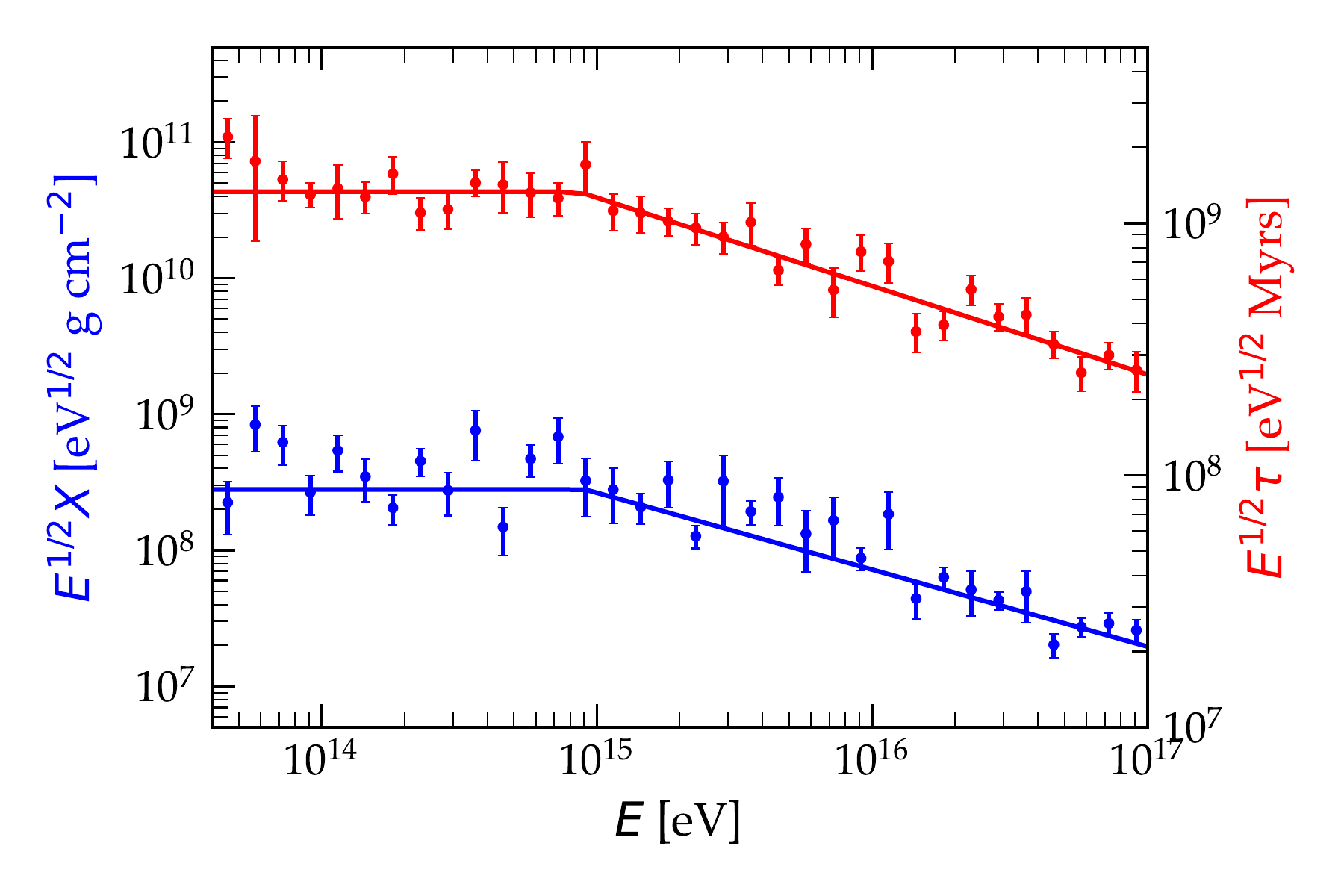}
    \caption{Cosmic ray escape time (red) and grammage (blue) as function of particle energy for a simulation box with $H = 2$ kpc, $h = 50$ pc, $\eta = 0.5$, $\lambda_c \approx 30$ pc and $\mu = 0$. Scatter points represent average time (grammage) over energy bin from simulation data and their best fit assuming a broken power-law is shown in red (blue) solid line. Both escape time and grammage are scaled by $E^{1/2}$ to facilitate the visualization of the transition between diffusion and drifts.}
    \label{fig:grammage_time_base}
\end{figure}

This simple picture is qualitatively confirmed by the results of our simulations: the resulting escape time and grammage (averaged over particle energy bins) are illustrated in Fig.~\ref{fig:grammage_time_base} for a case with $\eta=0.5$. The weak modulation visible in these results, as well as in the analogous results presented below, is a numerical artifact arising from binning the simulated particles in energy and averaging the relevant quantities within each bin. The adopted binning represents a compromise between reducing these statistical fluctuations and retaining sufficient energy resolution around the diffusion--drift transition.
\par Simulations show the appearance of a break in the confinement time (and the grammage) at $\sim1$ PeV associated with the emergence of drifts, as demonstrated by the fact that the energy dependence of the escape time in that region is $\tau \propto E^{-1}$. 
We fit both simulation data with a broken-power law of the form 
\begin{equation}
    \label{eq:BPL}
    \tau(E)\propto\left(\frac{E}{E_0}\right)^{-\gamma_1} \left(1 + \left(\frac{E}{E_b}\right)^{1/\omega}\right)^{-(\gamma_2 - \gamma_1) \omega}
\end{equation}
and find changes of spectral index $\Delta \gamma\approx0.3-0.7$, which are in good agreement with a model in which the \emph{knee} is to be attributed to the diffusion-drift transition.
However, as anticipated above, in this model the CR grammage is problematic: measurements of the B/C ratio require that $\rchi(1 \, \rm TeV)\approx0.6$ g cm$^{-2}$~\cite{Schroer2021}, lower than the grammage resulting from our simulations by about two orders of magnitude, with $\rchi(100 \, \rm TeV)\approx30$ g cm$^{-2}$. The grammage becomes of the order $10^{-2}$ g cm$^{-2}$ only after the diffusion-drift transition, at $E\sim10$ PeV. This important constraint is mainly due to the very inefficient perpendicular diffusion, that forces particles to remain in the disc region for too long a time~\cite{Giacinti2018}. This finding makes a model of the \emph{knee} based on the transition from perpendicular diffusion to drifts highly unlikely if not unfeasible.

\subsection{Galactic CR escape in the presence of $z$-component of the magnetic field ($\mu\neq0$)}
\label{sec:results_C}

These problems can be partially alleviated by introducing a $z$-component of the magnetic field, which activates escape through parallel diffusion. Such a process requires the particles to cover a distance much larger than $H$ to escape the mock Galaxy, but $D_\parallel\gg D_\perp$, hence this class of models may be of interest. 

\begin{figure}[t]
    \centering
    \includegraphics[width=0.497\linewidth]{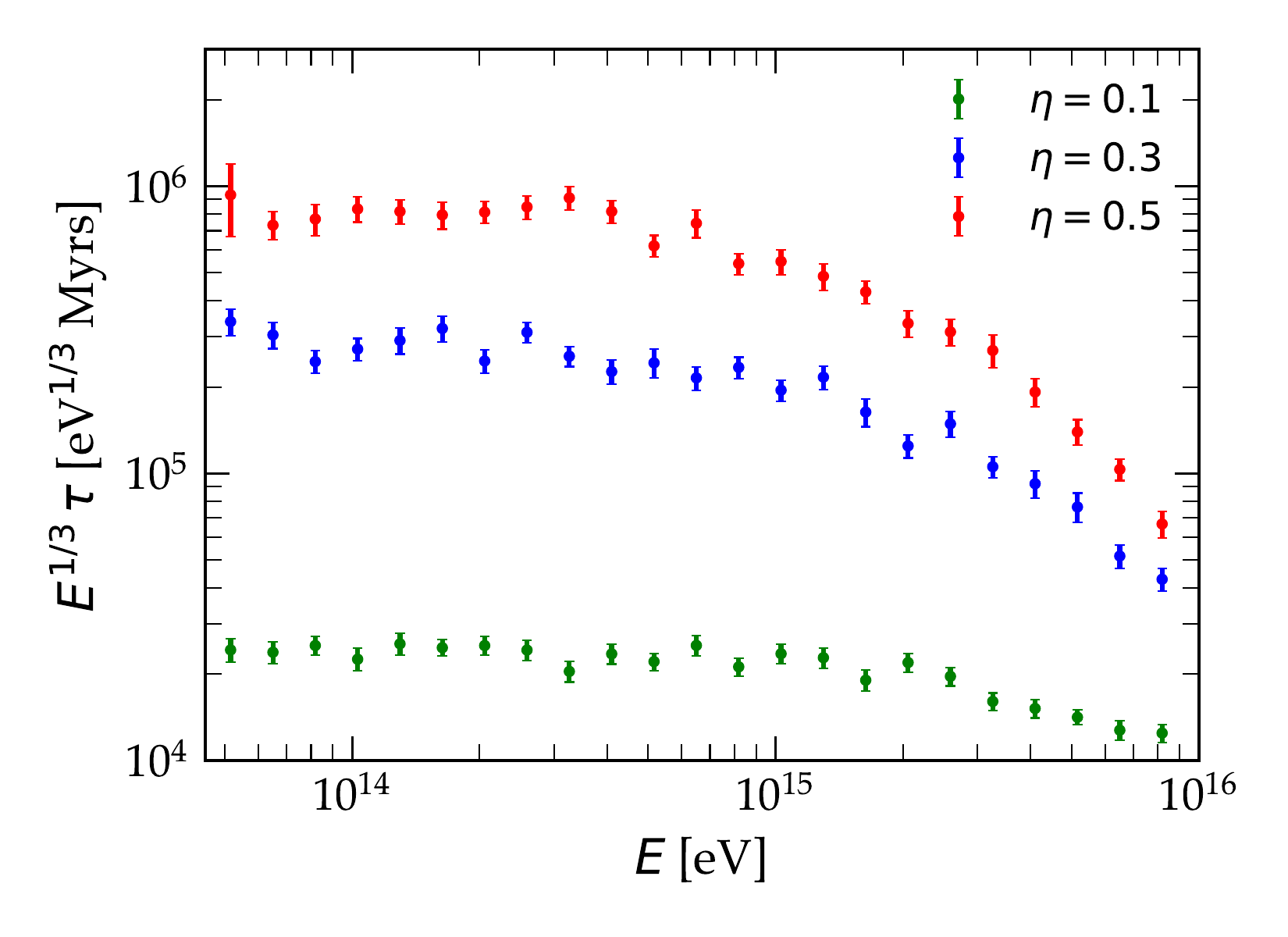}
    \includegraphics[width=0.497\linewidth]{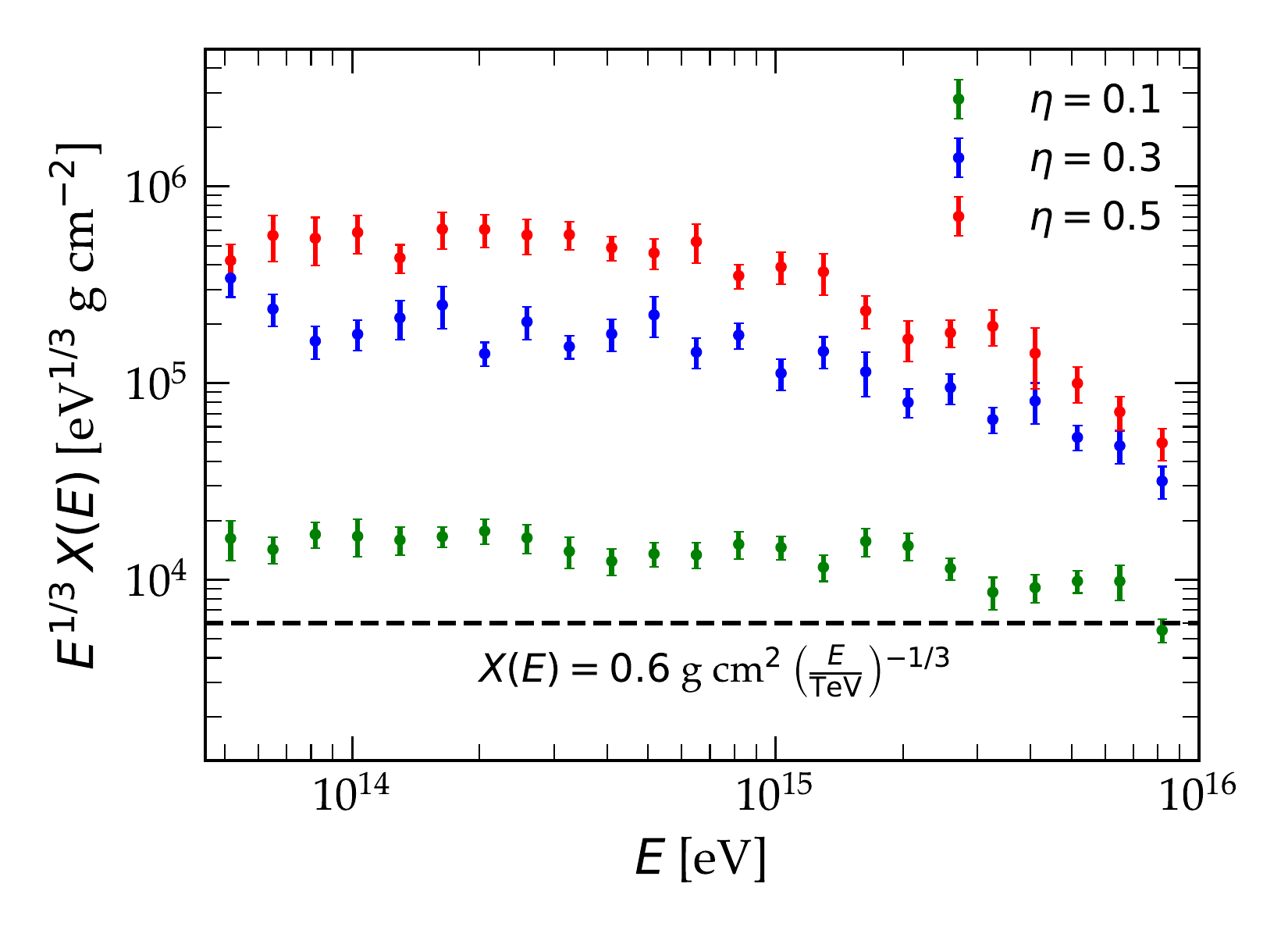}
    \caption{Left panel: Corresponding escape times as function of particle energy for simulations with $H = 2$ kpc, $h = 50$ pc, $\lambda_c \approx 30$ pc, $\mu = 0.1$ and $\eta=0.1, 0.3, 0.5$ (green, blue and red scatter points). Values scaled by $E^{1/3}$ to facilitate the visualization of the transition between diffusion and drifts. Right panel: Cosmic ray grammage as function of particle energy from the same simulations. Black dashed line denotes the predicted grammage from the extrapolation of the grammage observed at TeV (by B/C observations) to higher energies assuming $\rchi\propto E^{-1/3}$. Values also scaled by $E^{1/3}$.}
    \label{fig:grammage_enhanced}
\end{figure}

\begin{figure}[t]
    \centering
    \includegraphics[width=0.497\linewidth]{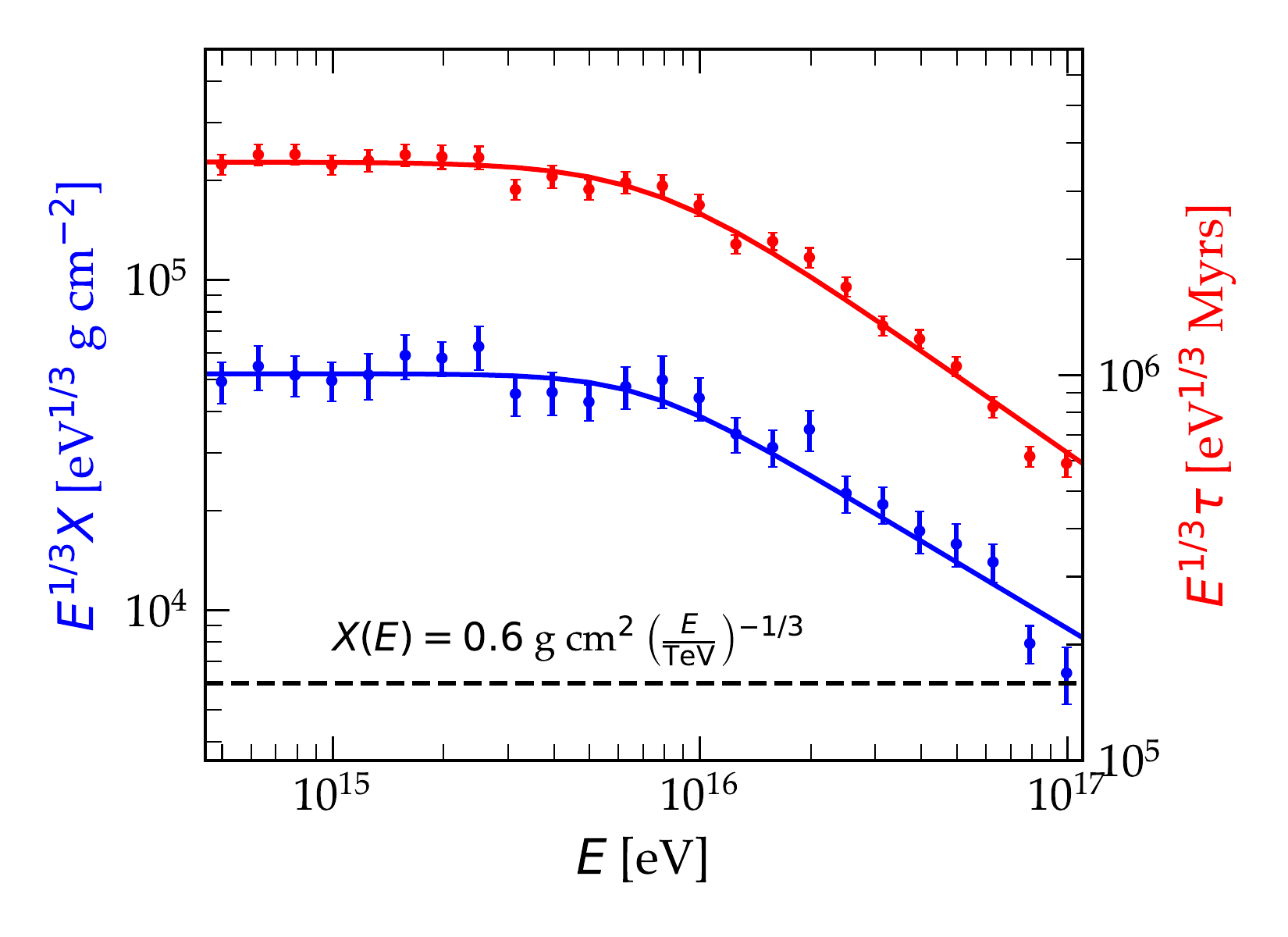}
    \includegraphics[width=0.497\linewidth]{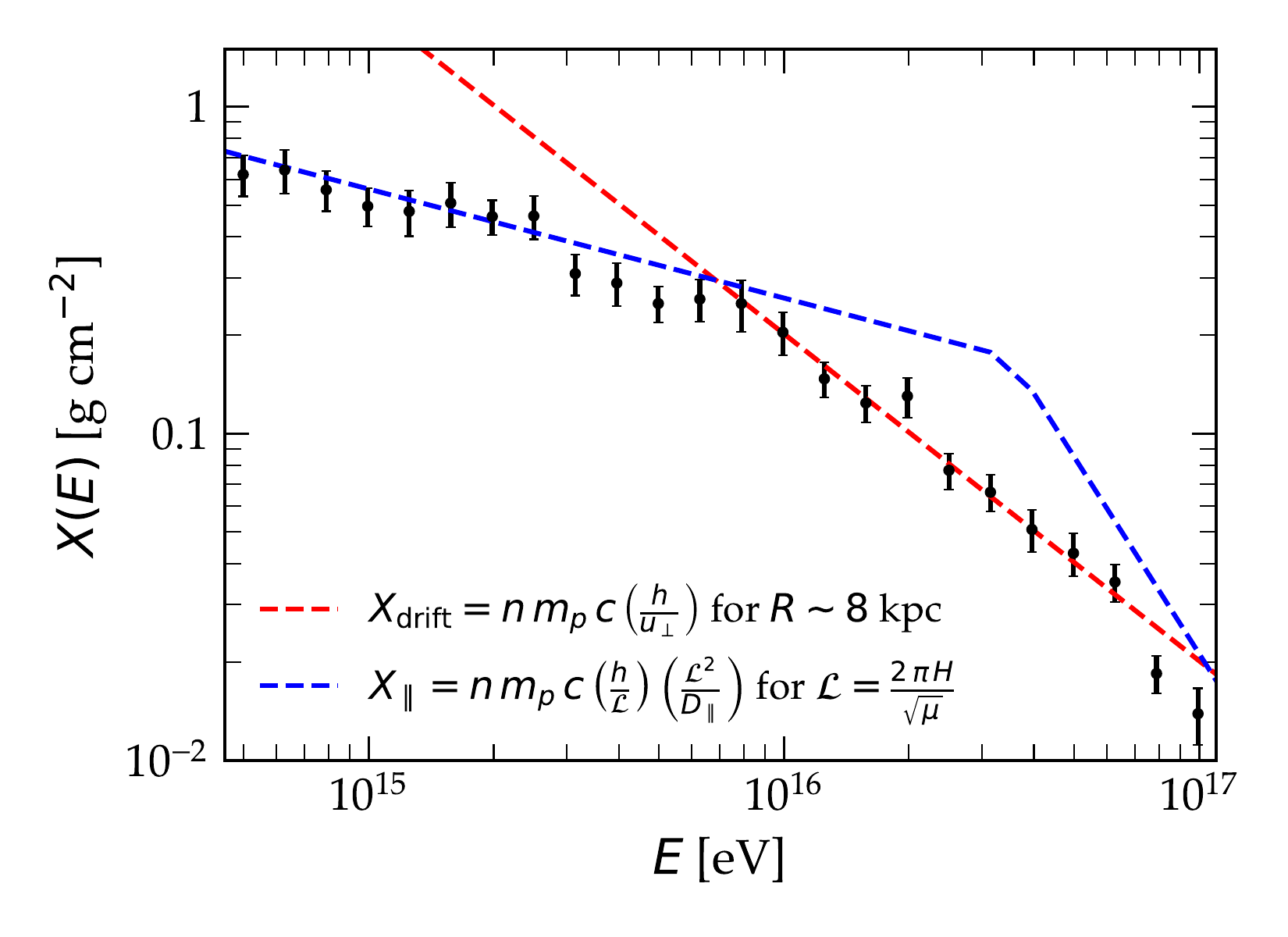}
    \caption{Left panel: Cosmic ray escape time (red) and grammage (blue) as function of particle energy for a simulation box with $H = 2$ kpc, $h = 50$ pc, $\eta = 0.7$, $\lambda_c \approx 30$ pc, $\mu = 0.2$. Scatter points represent average time (grammage) over energy bin from simulation data and their best fit assuming a broken power-law is shown in red (blue) solid line. Both escape time and grammage are scaled by $E^{1/3}$ to facilitate the visualization of the transition between diffusion and drifts. Black dashed line denotes the predicted grammage from the extrapolation of the grammage observed at TeV (by B/C observations) to higher energies assuming a $\propto E^{-1/3}$ scaling. Right panel: Cosmic ray grammage (black scatter points) as function of particle energy from the same simulation setup. Predicted grammage via parallel diffusion and drifts (computed from Eqs. \ref{eq:grammage_parallel_general} and \ref{eq:grammage_drift_general}) shown in blue and red dashed lines, respectively.}
    \label{fig:grammage_time_enhanced_highturb}
\end{figure}

Our results are presented in Fig.~\ref{fig:grammage_enhanced} for a small vertical component ($\mu=0.1$ in Eq.~\ref{eq:coherent_field}) and several values of the turbulence parameter $\eta$. At low energies one can clearly appreciate that the energy dependence of the escape time and the grammage are $\propto E^{-1/3}$, as expected for parallel diffusion. A transition to transport affected by drifts can be seen in all cases considered, but a clear transition only occurs for high values of $\eta$, for instance $\eta=0.5$. However, the right panel of Fig.~\ref{fig:grammage_enhanced} shows that such scenarios also lead to too large a grammage in the $\sim$ TeV energy range, and are therefore disfavored for the same reasons discussed above in the scenario in which perpendicular diffusion competes with drifts. 
We also explored the effect of increasing the vertical component of the GMF, performing simulations with $\mu=0.2$. In this field, a clear spectral break is found in the resulting escape time and grammage for simulations with high turbulence (shown in left panel of Fig.~\ref{fig:grammage_time_enhanced_highturb} for $\eta=0.7$). The grammage is lower than found in simulations with $\mu=0.1$, but remains larger than the measured values in the TeV energy range by about one order of magnitude.  

This rich phenomenology can be easily understood following the same line of reasoning as outlined in the case of a purely azimuthal field. 
Escape through diffusion parallel to field lines requires that particles cover diffusively a distance of the order of 
\begin{equation}
    L_\parallel \approx 2\pi R_\odot\left(\frac{H}{\sqrt{\mu}R_\odot}\right),
\end{equation}
where $R_\odot\sim8$ kpc is the radial position of the Sun in the Galaxy and $\frac{H}{\sqrt{\mu}R_\odot}$ is a good estimate of the number of twirls that the field line must perform before reaching the edge of the halo (escape). Therefore, the grammage accumulated by low-energy particles (for which the effect of drifts is still subdominant and parallel diffusion is the dominant escape mechanism) is given by
\begin{equation}
    \label{eq:grammage_parallel_general}
    \rchi_\parallel(E) = n_{\rm ISM} \, m_p\frac{ c \,  h \, L_\parallel}{D_\parallel} = \frac{n_{\rm ISM} \, m_p \, h \, 6 \pi \, H \, \eta^2}{\sqrt{\mu} \, r_L^{1/3} \, \lambda_c^{2/3}},
\end{equation}
for a number density of the ISM $n_{\rm ISM}$, proton mass $m_p$ and parallel diffusion coefficient $D_\parallel \approx \frac{c \, r_L^{1/3} \, \lambda_c^{2/3}}{3 \eta^2}$. On the other hand, the grammage accumulated by drifts for high-energy particles can be estimated as
\begin{equation}
    \label{eq:grammage_drift_general}
    \rchi_{\rm drift}(E) = n \, m_p \, c\left(\frac{h}{u_\perp}\right) = \frac{2 \, n \, m_p \, h \, R}{r_L},
\end{equation}
assuming that drifts are originated only from field-line curvature, so that $u_\perp=\frac{c \, r_L \, }{2 R}$ for a radius to the Galactic center $R=R_\odot$. 

\begin{figure}[t]
    \centering
    \includegraphics[width=0.55\linewidth]{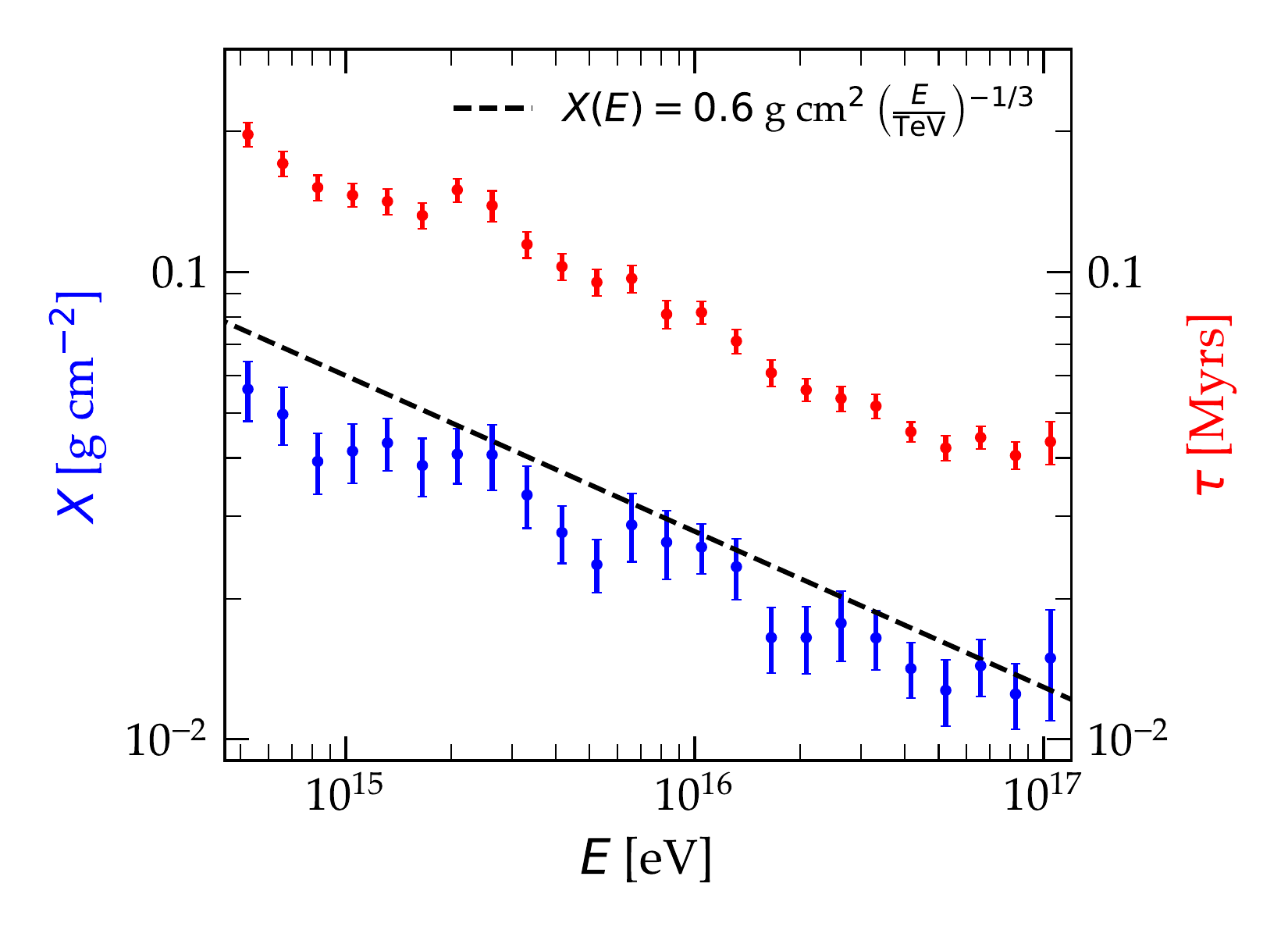}
    \caption{Cosmic ray escape time (red) and grammage (blue) as function of particle energy for a simulation box with $H = 2$ kpc, $h = 50$ pc, $\eta = 0.2$, $\lambda_c \approx 30$ pc, $\mu = 0.2$. Scatter points represent average time (grammage) over energy bin from simulation data. Black dashed line denotes the predicted grammage from the extrapolation of the grammage observed at TeV (by B/C observations) to higher energies assuming a $\propto E^{-1/3}$ scaling.}
    \label{fig:grammage_time_ehanced_lowturb}
\end{figure}

For the parameter values adopted in the simulation of Fig.~\ref{fig:grammage_time_enhanced_highturb} ($H = 2$ kpc, $h = 50$ pc, $\eta = 0.7$, $\lambda_c \approx 30$ pc, $\mu = 0.2$), we can estimate the grammage using Eqs.~\ref{eq:grammage_parallel_general} and \ref{eq:grammage_drift_general}. As one can see in the right panel of Fig.~\ref{fig:grammage_time_enhanced_highturb}, we find a good agreement between estimates (dashed lines) and results (scatter points). Based on Eq.~\ref{eq:grammage_parallel_general}, we also find the magnetic turbulence level which corresponds to a grammage $\rchi\sim 0.6$ g cm$^{-2}$ at 1 TeV. We find that $\eta=0.2$ is expected to produce a comparable value at this energy, with
\begin{equation}
\label{eq:grammage_parallel}
    \rchi(E = 1 \, \rm{TeV}) = 0.43 \, g \, cm^{-2} \left(\frac{n}{0.5 \, cm^{-3}}\right)\left(\frac{h}{50 \, pc}\right)\left(\frac{H}{2 \, kpc}\right) \left(\frac{\eta}{0.2}\right)^2 \left(\frac{0.2}{\mu}\right)^{1/2}\left(\frac{30 \, pc}{\lambda_c}\right)^{2/3}.
\end{equation}
After running simulations with this turbulence level, we find a CR grammage (blue scatter points in Fig.~\ref{fig:grammage_time_ehanced_lowturb}) in very good agreement with the expected value (black dashed line), but no sign of a break associated to the diffusion-drift transition. Introducing the field parameters of these simulations into Eq.~\ref{eq:grammage_drift_general} for $R=8$ kpc, we estimate the grammage via drifts at 1 PeV as
\begin{equation}
\label{eq:grammage_drift}
    \rchi(E = 4 \, \text{PeV}) = 0.49 \, \text{g cm}^{-2} \, \left(\frac{n}{0.5 \, \rm cm^{-3}}\right)\left(\frac{h}{50 \, \rm pc}\right)\left(\frac{R}{8 \, \rm kpc}\right),
\end{equation}
which is similar to the value found at 1 TeV in Eq.~\ref{eq:grammage_parallel}. Drifts are now a significantly less efficient propagation mechanism in comparison with parallel diffusion. Therefore, reproduction of the grammage at low-energies via parallel diffusion implies the disappearance of a transition to drifts, since parallel diffusion becomes totally dominant in the TeV-PeV energy range.

\subsection{The case of energy-independent cosmic ray parallel diffusion}
\label{sec:results_D}

The main difficulty in associating the \emph{knee} with a
transition from diffusive to drift-dominated escape is the energy dependence
of the diffusive escape time. Because secondary-to-primary ratios have not
been measured above $\sim$TeV, this energy dependence is not directly
constrained in the range relevant to the \emph{knee}. Both theoretical and
observational considerations therefore motivate exploring a nearly
energy-independent $D_\parallel$ at high energy. In particular, the
flattening reported in the B/C data from DAMPE~\cite{Dampe2022} and
NUCLEON~\cite{Grebenyuk2019} has been interpreted in terms of
energy-independent CR transport~\cite{Recchia2024}, although accumulation of
grammage near the sources provides an alternative explanation
\cite{Ambrosone2025,Yang2025}.

\par It is important, however, to distinguish two
physically different mechanisms that can produce this behavior. The first is
a resonant-scattering scenario: a $k^{-2}$ spectrum extends over a broad
range of wavenumbers and retains power at the resonant scale. For a
one-dimensional spectrum $P(k)\propto k^{-\alpha}$, quasi-linear theory gives
$D_\parallel\propto r_L^{2-\alpha}$, so that $\alpha=2$ yields
$D_\parallel\propto r_L^0$. We use this spectrum as a phenomenological
representation of the parallel scaling associated with the anisotropic
Alfv\'enic cascade of Goldreich and Sridhar~\cite{Gold1994,Gold1995}, without
attempting to reproduce its anisotropy or dynamics.

\par The second mechanism is
motivated by MHD intermittency, in which scattering may be dominated by
occasional encounters with large magnetic structures on scales comparable
to the Larmor radius~\cite{Lemoine2023,Kempski2023,Kempski2025}. The energy
dependence of diffusion in this case remains uncertain, although some MHD
simulations find an approximately energy-independent coefficient
\cite{Kempski2025}. Our random-phase synthetic turbulence does not contain
the non-Gaussian structures required to model intermittency directly. We
instead mimic the limiting transport phenomenology by concentrating the
turbulent power in a narrow range around the coherence scale, idealized as
$W(k)\propto\delta(k-1/\lambda_c)$, following a strategy similar to that of
Ref.~\cite{Pezzi2024}. For $r_L\ll\lambda_c$, this spectrum contains
essentially no resonant power, and particles are redirected as they follow
magnetic-field lines whose direction changes on the coherence scale. The
narrow-band construction should therefore be regarded as an idealized proxy
for this transport phenomenology, not as a realization of MHD intermittency. In both scenarios, we assume that transport below $\sim$TeV is controlled by
other processes, such as scattering-off fluctuations self-generated by
CRs~\cite{Farmer2004,Blasi2012}.

\par
\begin{figure}[t]
    \centering
    \includegraphics[width=0.49\linewidth]{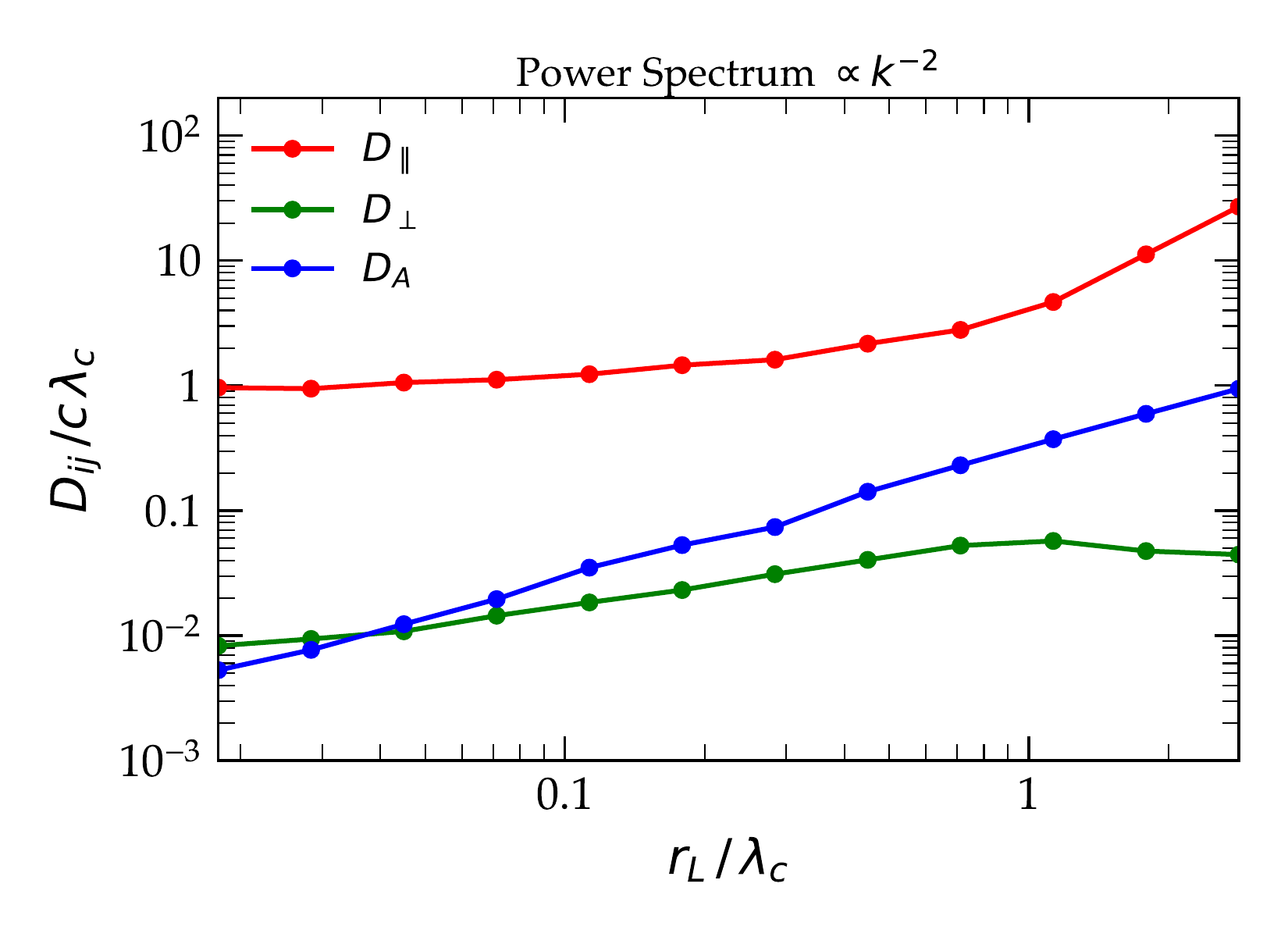}
    \includegraphics[width=0.49\linewidth]{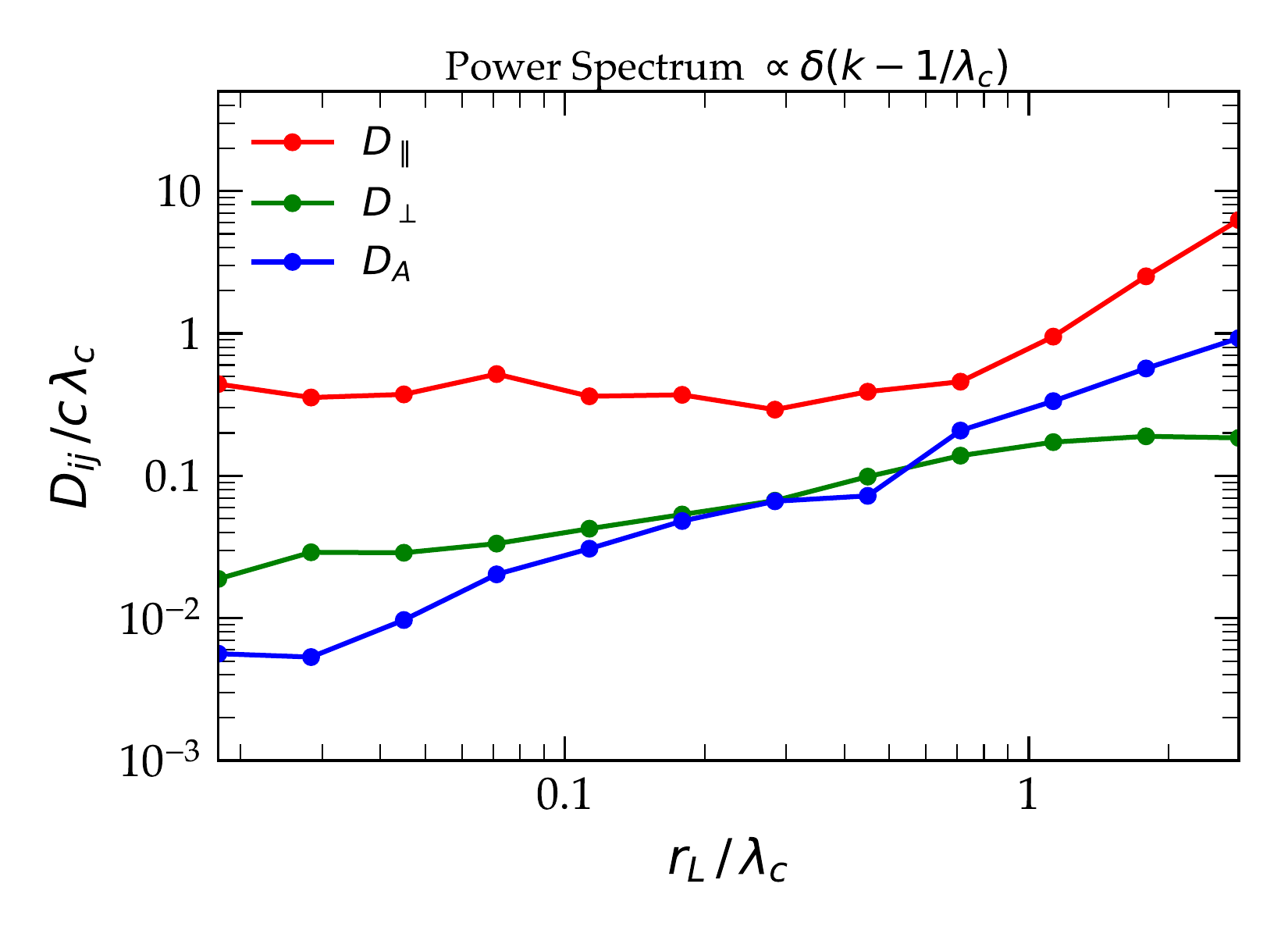}
    \caption{Parallel (red), perpendicular (green) and antisymmetric (blue) coefficients of the diffusion tensor (Eq.~\ref{eq:diffusion_tensor}) as function of particle rigidity for turbulent spectra $W(k)\propto k^{-2}$ with $\delta B / B_0 = 0.7$ (left panel) and $W(k)\propto \delta(k\sim1/\lambda_c)$ with $\delta B / B_0 = 2.0$ (right panel). Scatter points represent average diffusion coefficient over rigidity bin. Coefficients normalized over $c\lambda_c$ and particle rigidity represented by ratio $r_L(E)/\lambda_c$.}
    \label{fig:flat_diffusion_coefficients}
\end{figure}

 We analyze the narrow-band case first by
setting $L_{\text{min}}=\lambda_c/2$ and $L_{\text{max}}=2\lambda_c$. Following
the procedure used for Fig.~\ref{fig:coefficients}, we calculate the
diffusion coefficients in a coherent field directed uniformly along the
$x$-axis. The right panel of Fig.~\ref{fig:flat_diffusion_coefficients} shows
$D_{\parallel}$ (red), $D_{\perp}$ (green), and $D_A$ (blue) for
$\delta B/B_0=2.0$. At low rigidity, $D_{\parallel}$ is approximately energy
independent, $D_{\perp}$ retains the qualitative behavior discussed in
Sec.~\ref{sec:results_A}, and $D_A$ increases approximately linearly with
rigidity.

\begin{figure}[t]
    \centering
    \includegraphics[width=0.497\linewidth]{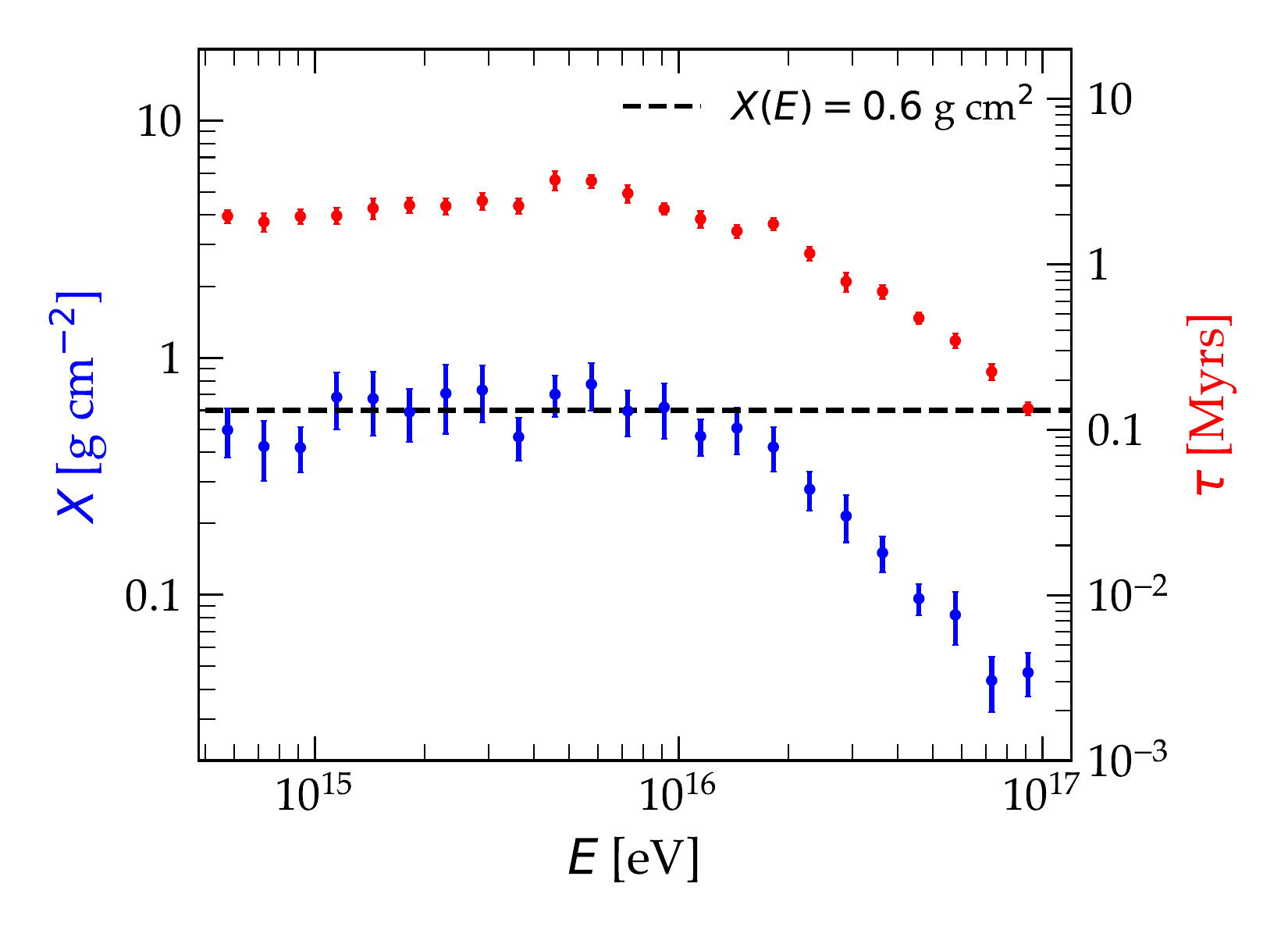}
    \includegraphics[width=0.497\linewidth]{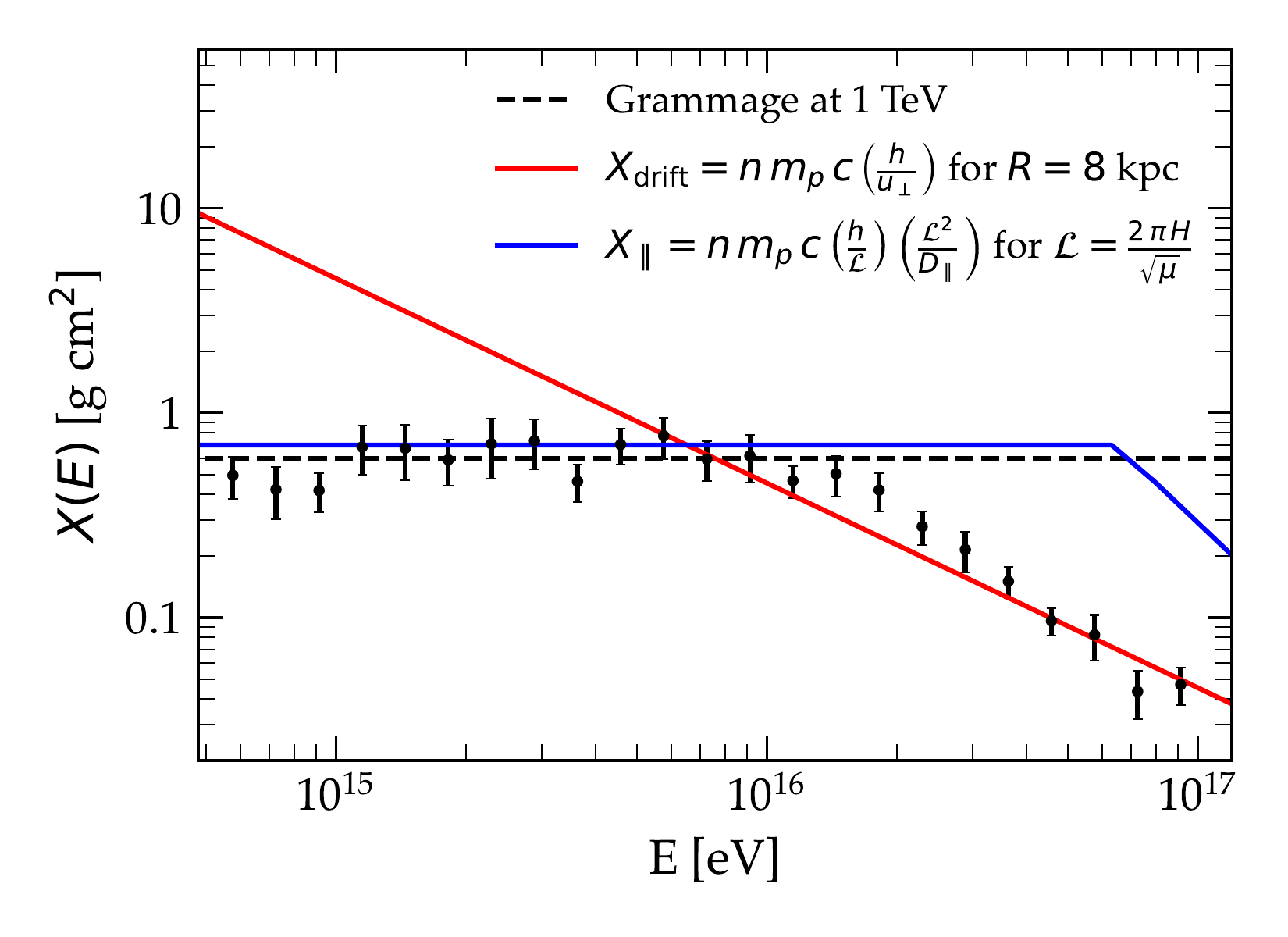}
    \caption{Cosmic ray escape time (red) and grammage (blue) as function of particle energy for a simulation box with $H = 2$ kpc, $h = 50$ pc, $\eta = 2.0$, $\lambda_c \approx 30$ pc, $\mu = 0.2$. In this case, the spectrum of the synthetic turbulent magnetic field was constructed resembling a Dirac delta function $\delta(k\sim1/\lambda_c)$ in order to make the diffusion coefficient energy-independent in the TeV-PeV range, following the work in~\cite{Pezzi2024}. Scatter points represent average time (grammage) over energy bin from simulation data. Black dashed line denotes the grammage observed at TeV by B/C observations.}
    \label{fig:flat_diffusion_grammage2}
\end{figure}

The escape times and grammage that we computed in this case are displayed in the left panel of Fig.~\ref{fig:flat_diffusion_grammage2} as red and blue scatter points, respectively. As expected, both quantities turn out to be energy independent up to $\sim10$ PeV, while a break occurs due to the onset of drifts. Moreover, the grammage at low energies ($E<10$ PeV) is in very good agreement with the value inferred at 1 TeV from direct CR observations. The grammage at high energies ($E>10$ PeV) follows the expected energy dependence related to drifts in the left panel of Fig.~\ref{fig:flat_diffusion_grammage2}. There, we compare the grammage of our simulations (black scatter points) to the predicted grammage by Eqs. \ref{eq:grammage_parallel_general} and \ref{eq:grammage_drift_general}, finding a good agreement in both regimes. Notice however that this scenario requires strong turbulence, $\eta=2$: it is interesting that even with such strong turbulence level, the drifts still affect the results (one would naively expect drifts to be suppressed by turbulence, but clearly this is a gradual effect). 

\begin{figure}[t]
    \centering
    \includegraphics[width=0.497\linewidth]{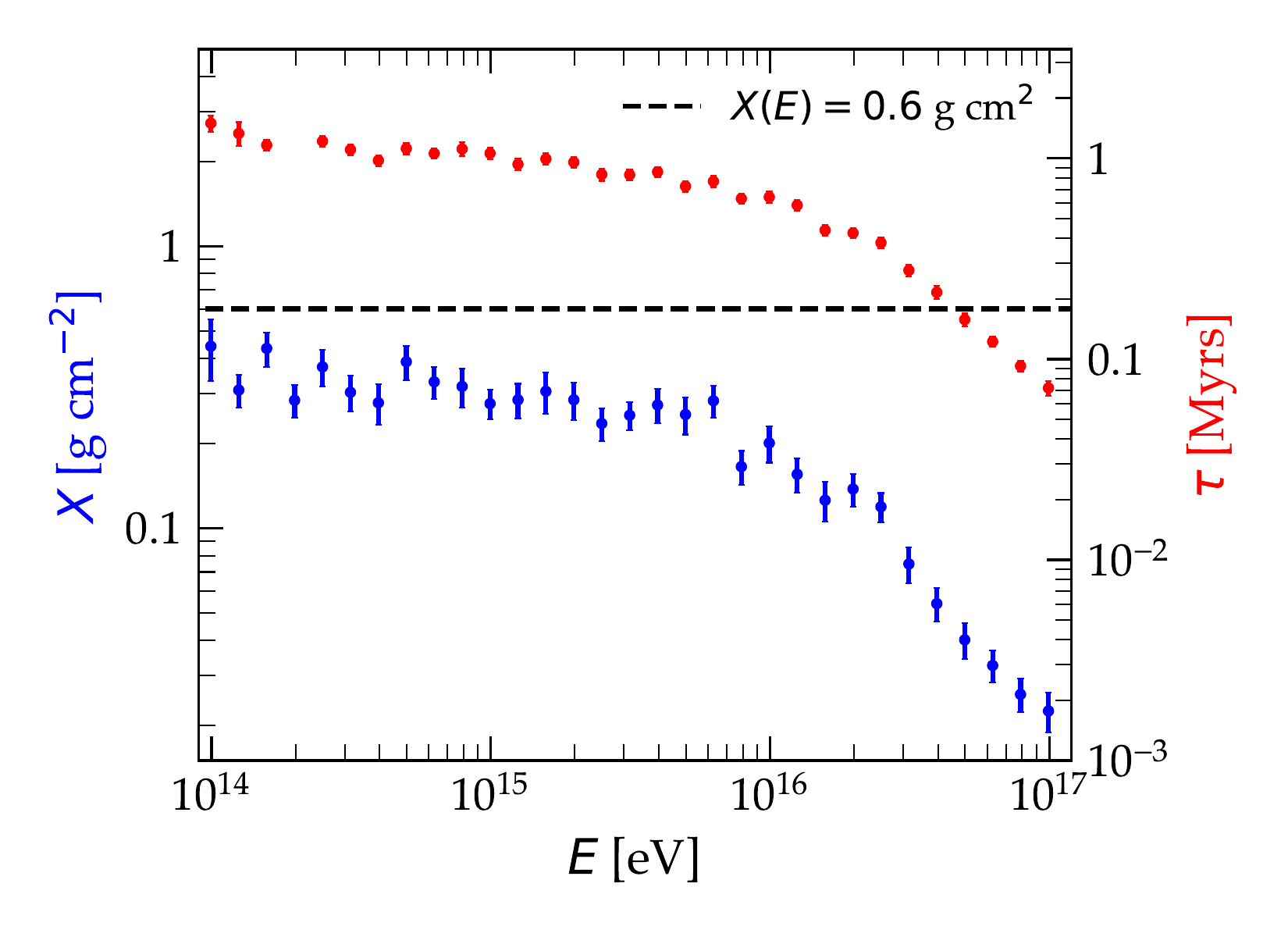}
    \includegraphics[width=0.497\linewidth]{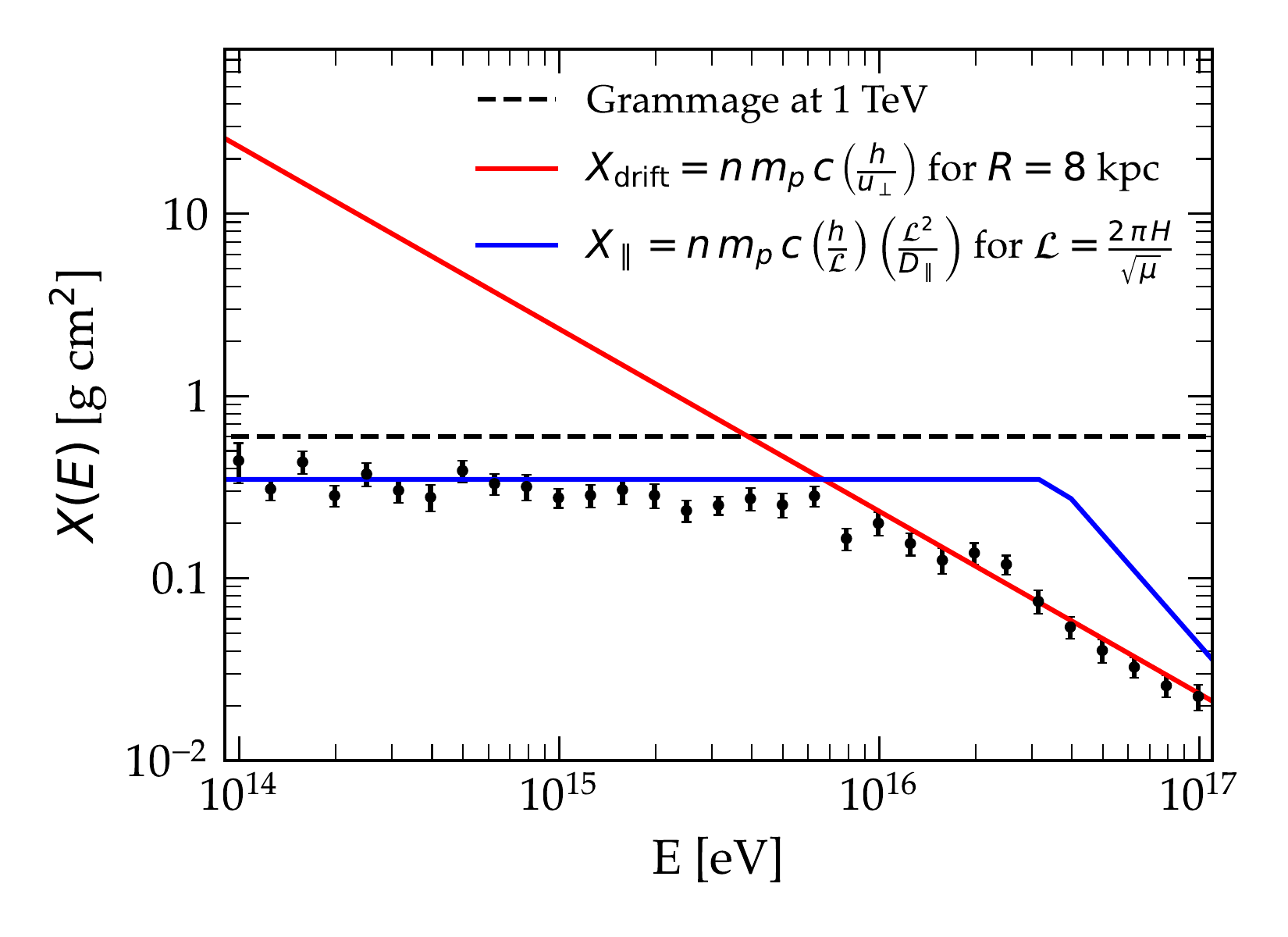}
    \caption{Cosmic ray escape time (red) and grammage (blue) as function of particle energy for a simulation box with $H = 2$ kpc, $h = 50$ pc, $\eta = 0.7$, $\lambda_c \approx 30$ pc, $\mu = 0.2$. In this case, the spectrum of the synthetic turbulent magnetic field was constructed with index $s=2$ in order to make the diffusion coefficient energy-independent in the TeV-PeV range. Scatter points represent average time (grammage) over energy bin from simulation data. Black dashed line denotes the grammage observed at TeV by B/C observations.}
    \label{fig:flat_diffusion_grammage}
\end{figure}

An energy-independent cosmic ray diffusion can also be obtained by assuming a turbulence spectrum $\propto k^{-2}$. We modify the turbulence spectrum in our numerical setup accordingly (setting $s=2$ in Eq.~\ref{eq:spectrum}) and repeat our test particle simulations, using $\eta = 0.7$, $\lambda_c \approx 30$ pc and $\mu = 0.2$. 
The corresponding diffusion coefficients are shown in the
left panel of Fig.~\ref{fig:flat_diffusion_coefficients}. As expected for
$\alpha=2$, $D_{\parallel}$ is approximately energy independent at low
rigidity. The qualitative behavior of $D_{\perp}$ is similar to that of the
narrow-band case, while $D_A$ again increases approximately linearly with
rigidity.
The particle escape times obtained in this case are shown in the left panel of Fig.~\ref{fig:flat_diffusion_grammage} as red scatter points, while the CR grammage is shown with blue scatter points. The observational value of the grammage at 1 TeV is displayed with a black dashed line. In both measurements, the results are similar to those of Fig.~\ref{fig:flat_diffusion_grammage2} and follow an approximately constant value up to $\sim10$ PeV, where a spectral break is observed. Contrary to the scenario of Fig.~\ref{fig:flat_diffusion_grammage2}, here a low turbulence level ($\eta<1$) is sufficient to produce energy-independent diffusion at sub-PeV energies. The break here can again be attributed to the onset of drifts, since the grammage at energies above 10 PeV is in good agreement with the predicted grammage from  Eq.~\ref{eq:grammage_drift_general}, shown in the right panel of Fig.~\ref{fig:flat_diffusion_grammage}. Here, the grammage from the simulations (black scatter points) is displayed in the right panel of Fig.~\ref{fig:flat_diffusion_grammage}, along with the expected grammage due to parallel diffusion (blue line, from Eq.~\ref{eq:grammage_parallel_general}) and drifts (red line). Again, the measured value of grammage at 1 TeV is shown as a dashed black line.

\section{Discussion and Conclusions}
\label{sec:conclusion}

We used test particle numerical simulations of CR propagation in homogeneous isotropic magnetic turbulence and different configurations of the regular field to test the possibility that the \emph{knee} may arise from the competition between diffusion and drifts. The latter can be induced by either gradients or curvature of the regular magnetic field, of the type that can be expected in the Galaxy. 

We first calculated the components of the diffusion tensor, including the antisymmetric ones, the so called Hall terms. This part of the calculations had a twofold purpose: on one hand, we established contact with previous literature, especially to confirm the anomalous behaviour of $D_\perp$ as a function of energy~\cite{DeMarco2007,Dundovic2020,Mertsch2025b}. On the other hand, we wanted to compute the Hall terms and confirm the linear energy dependence postulated in previous literature \cite{Candia2002,Candia2004}. 

The competition between diffusion and drifts was investigated in three models: 1) purely azimuthal large scale magnetic field with Kolmogorov-like turbulence; 2)  regular magnetic field with a dominant azimuthal structure and a small component in the $z$-direction, plus Kolmogorov-like turbulence; 3) regular magnetic field with a dominant azimuthal structure and a small component in the $z$-direction plus a turbulent field either concentrated on one scale or distributed in $k$-space as $k^{-2}$. 

In the first model, the competition between diffusion and drift indeed produces a knee-like feature at an energy that, for typical values of the parameters, can be in the PeV region. However, the required field configuration, is such that only perpendicular diffusion contributes to CR escape, and this reflects in an exceedingly long time spent by particles in the disc, where the grammage accumulated at low energies turns out to largely exceed the observed grammage. This model is therefore ruled out as a possible explanation of the knee. 

In the second model, we consider the possibility that the regular magnetic field may have a small $z$-component: in this case, the very fast parallel diffusion leads to particle escape from the halo in too short a time for drifts to play a relevant role. Hence, also in this second class of models, the drift cannot be invoked to explain the knee. Some previous studies (\cite{Cerri2017, Recchia2024}) have also explored instances in which the Galactic magnetic field configuration is such that field lines are mainly parallel to the Galactic plane in the disk, while in the Galactic halo the field lines acquire a significant vertical component. In such models, the interplay between diffusion and drifts becomes more complex to assess, because diffusion becomes a function of the location, and even more important, because drifts reflect the curvature of magnetic fields, that is clearly different from one location to another.
\par In the third situation that we consider, we assume that the turbulence spectrum to be added to the regular field having a small $z$-component is either in the form of a peaked spectrum around $\lambda_c$ or a spectrum $k^{-2}$. In the first case we assume a large turbulence level ($\eta=2$), while in the second case we adopt $\eta=0.7$. Both these configurations lead to a parallel mean free path that is roughly energy independent up to $\sim 10$ PV. This situation leads to an escape time and grammage that are in reasonable agreement with low energy observations, and lead to the appearance of a \emph{knee} around $\sim$ PeV energy, due to the competition between parallel transport and drifts. 

Our conclusion is that properly explaining the appearance of a \emph{knee} and the observed spectral steepening across the \emph{knee} requires a special concoction of particle transport and curvature properties of the large scale Galactic magnetic field. It cannot however be excluded that such a combination may be achieved in nature. It is important to emphasize that if such a situation really occurs then the implication is that the accelerators of Galactic CRs must be more effective than ever believed, in that the maximum energy is required to be much larger than the position of the knee. 

\begin{acknowledgments}
This work has been partially funded by the European Union - Next Generation EU under PRIN-MUR 2022TJW4EJ ``Unveiling the footprints of the cosmic ray journey through the Galaxy and beyond'' and under MUR National Innovation Ecosystem grant ECS00000041 - VITALITY/ASTRA - CUP D13C21000430001. 
\end{acknowledgments}

\section*{Data Availability}
Data produced in the analysis of this study are available upon reasonable request.

\bibliographystyle{myapsrev4-2}
\bibliography{2026-EspinosaCastro}

\end{document}